\documentclass[preprint2]{emulateapj}
\slugcomment{{\sc Accepted to ApJ:} August 8, 2012} 
\usepackage{epstopdf}
\usepackage{url}
\shorttitle{TEXES Observations Of NGC 7538 IRS9}
\shortauthors{Barentine and Lacy}

\begin{document}

\title{A Comparative Astrochemical Study Of The High-Mass Protostellar Objects NGC 7538 IRS 9 and IRS 1}

\author{John C. Barentine and John H. Lacy}
\affil{Department of Astronomy, The University of Texas at Austin}
\affil{1 University Station, C1400, Austin, TX 78712-0259 USA}

\begin{abstract}
We report the results of a spectroscopic study of the high-mass protostellar object NGC 7538 IRS 9 and compare our observations to 
published data on the nearby object NGC 7538 IRS 1.  Both objects originated in the same molecular cloud and appear to be at different 
points in their evolutionary histories, offering an unusual opportunity to study the temporal evolution of envelope chemistry in objects sharing 
a presumably identical starting composition.  Observations were made with the Texas Echelon Cross Echelle Spectrograph (TEXES), a 
sensitive, high spectral resolution ($R = {\lambda}/{\Delta}{\lambda}\simeq$ 100,000) mid-infrared grating 
spectrometer.  Forty-six individual lines in vibrational modes of the molecules C$_2$H$_2$, CH$_4$, HCN, NH$_3$ and CO were detected, 
including two isotopologues ($^{13}$CO, $^{12}$C$^{18}$O) and one combination mode (${\nu}_{4}+{\nu}_{5}$ C$_2$H$_2$).  Fitting 
synthetic spectra to the data yielded the Doppler shift, excitation temperature, Doppler $b$ parameter, column density and covering factor for 
each molecule observed; we also computed column density upper limits for lines and species not detected, such as HNCO and OCS.  We 
find differences among spectra of the two objects likely attributable to their differing radiation and thermal environments.  Temperatures and 
column densities for the two objects are generally consistent, while the larger line widths toward IRS 9 result in less saturated lines than 
those toward IRS 1.  Finally, we compute an upper limit on the size of the continuum-emitting region ($\sim$2000 AU) and use this constraint 
and our spectroscopy results to construct a schematic model of IRS 9.
\end{abstract}


\section{Introduction\label{sec:intro}}

Details of the pre-Main Sequence (pre-MS) life cycles of low-mass stars such as the Sun ($M$ $\sim$ 1$M_{\odot}$) have been gleaned 
from decades of research; see \citet{McKeeOstriker07} for a recent review.  This is thanks in part to the relative amenability of low-mass 
protostellar objects to study.  However, details of high-mass ($M$ $>$ 8$M_{\odot}$) star formation are not as well understood at present, 
in part because high-mass stars are not generally forming near the solar neighborhood and remain deeply embedded during their pre-MS 
evolution.   An important unresolved issue is whether high-mass stars form in a manner resembling a scaled-up version of low-mass  star 
formation \citep{KetoZhang10, Johnston11} or through other mechanisms uncharacteristic of the low-mass case such as collisions/mergers 
\citep{Bonnell98, Bonnell05, BallyZinnecker05} and competitive accretion in clusters  \citep{Bonnell01, Bonnell04}.  Clearly, 
the environments in which high-mass star formation takes place seem to influence the formation process itself.  Massive stars are most often 
seen to form in large groups (``OB associations'') near the edges of dense clumps in giant molecular clouds (GMCs; \citealt{YZ07}).  \citet 
{ElmegreenLada77} first suggested that spatially distinct subgroups of stars within OB associations trigger waves of self-propagating star 
formation in GMCs via shock/ionization fronts that progress from the outside in.  This process may account for observations of OB subgroups 
that lie in linear arrangements along the Galactic plane, showing a monotonically-increasing sequence of ages \citep{Blaauw64,Blaauw91}.
Rapid improvements in both observational capabilities and theoretical sophistication in recent years are advancing our knowledge of the 
circumstances of massive star formation and enabling these fundamental questions to be addressed.

The origin of these stars is of particular interest due to their role in the Galactic ecosystem.  In addition to serving as the principal sites of 
heavy element nucleosynthesis, the violent end of massive stars in supernovae injects considerable mechanical energy into the 
interstellar medium (ISM) while enriching it with metals \citep{Arnett96}.  The debris are reconstituted in molecular clouds from which new 
stars form, often accompanied by circumstellar disks \citep{Jiang08}.  These disks are evidently the source of accretion onto high-mass 
stars as they approach the MS in the Hertzsprung-Russell diagram and ignite hydrogen burning, then continue to accrete their final 
masses, moving parallel to the MS \citep{YZ07}.  Relatively little is known about the structure and chemical composition of massive 
protostellar envelopes, and even less about how they evolve in time.  Ready comparison with envelope models for low-mass objects is 
instructive but only of limited use, in part due to the radically different radiation environment in the high-mass case.  The composition and 
structure of the envelope are of interest as they determine the final mass of the star through an accretion mechanism whose details are still 
unclear.  Piecing together the components of protostellar envelopes is complicated by the observation that most objects are deeply 
embedded, particularly early in the formation process.   Spectroscopy of molecules is useful in delineating physical structures at high 
extinction where other methods are limited.  Pure rotational transitions of molecules in the millimeter probe a sufficiently low energy regime 
but such observations are typically of low spatial resolution.  Better resolution is achieved in the infrared where many ro-vibrational 
molecular lines are located.

We present high resolution, mid-infrared spectra of a number of molecules toward an embedded infrared source in the NGC 7538 
starforming complex, NGC 7538 IRS 9, and compare the results to a previous study of the nearby source NGC 7538 IRS 1 \citep{Claudia}.   
Both sources are presumed to harbor high-mass protostellar objects in their interiors.  Their proximity to one another suggests that, having 
formed within the same molecular cloud, the starting chemical abundances for each object should have been substantially similar.  A 
comparison of their current states provides insights about differences in their ages and/or divergence in their individual evolutionary 
histories.  

Weak free-free emission ($\leq$ 60 $\mu$Jy at 3.6 cm; \citealt{Sandell05}) is observed toward IRS 9.  Further evidence for the 
comparatively primitive nature of the envelope of IRS 9 is given by the detection of various ices in its \it Infrared Space Observatory \rm (ISO) 
spectrum (\citealt{Whittet96}; \citealt{Gibb04}).  These ices are not seen toward IRS 1, which appears to be in a relatively advanced point in 
its pre-MS evolution.  It is a considerably stronger source at centimeter wavelengths and shows evidence of both ionized and molecular 
components in a bipolar outflow originating in a disk seen nearly edge-on \citep{Sandell09, Qiu11}.  IRS 1 also shows signs of a 
circumstellar disk and an emerging compact \ion{H}{2} region; it is thought to be a forming late O star undergoing active accretion, 
suggesting a more advanced evolutionary state \citep {Lacy08, Beuther12}.  \citet{Sandell09} further note that the system is heavily accreting 
($\dot{M}$ $\sim$ 2 $\times$ 10$^{-4}$ $M_{\odot}$ yr$^{-1}$) and that the accretion may be episodic in nature.  On the basis of GHz line 
observations, \citet{Surcis11} conclude that the accretion  appears to be the result of radial infall from a ``torus'' thousands of Astronomical Units (AU) in size, traced by H$_2$O and CH$_3$OH masers, instead of the Keplerian disk favored by \citet{Pestalozzi04}.

Until recently, estimates of the distance to NGC 7538 placed it at $\sim$3.0 kpc \citep{HCS01, Brunt03, FR03, Balog04, RW05, Kameya06, 
Araya07}.  However, we adopted the distance of \citet{Moscadelli09}, who find a value of 2.65$^{+0.12}_{-0.11}$ kpc from trigonometric 
parallax measurements.  The total luminosity of IRS 9 is about 3.5 $\times$ 10$^{4}$ L$_{\odot}$ (\citealt{Sandell05}, corrected to the 
Moscadelli et al. distance) and it is a bright IRAS 12 $\mu$m source at $\sim$60 Jy \citep{IRAS}.  A zero-age Main Sequence (ZAMS) star 
with this luminosity would have a spectral type of B0.5, an effective temperature $T_{eff}$ = 26200 K \citep{Panagia73}, and a mass of $\sim
$13 M$_\odot$ \citep{Ekstrom12}.  \citet{BoogertBlake04} suggest the existence of an inner, warm molecular emission region with a radius 
of $\sim$70 AU; the dust temperature a tthis distance for an object of IRS 9's luminosity would be $\sim$1000 K depending on the dust mass 
opacity.   Millimeter data show that IRS 9 is a site of active, massive star formation \citep{Sandell05} in which young stellar objects (YSOs) 
are driving a set of bipolar molecular outflows on a dynamical timescale of $\leq$ 20,000 years.  At least one of these outflows appears to be 
highly energetic.  \citet{Mitchell91} report the discovery of an outflow in IRS 9 with a velocity of 110 km s$^{-1}$, which they interpret to have 
emerged as recently as 1,200 years ago, perhaps indicating the beginning of a recent episode of accretion.  

Observations of these protostellar objects are  broadly consistent with the \citet{ElmegreenLada77} picture of sequential high-mass star 
formation, as elaborated on by others in the particular case of NGC 7538 \citep{Werner79, Fischer80, Dickel81}.  The \citet
{CampbellThompson84} model of NGC 7538 holds that a first generation of massive stars caused the visible \ion{H}{2} region, which 
expanded and impinged on the neighboring molecular cloud; this compressed the cloud and generated shocks that caused the observed 
near-infrared fluorescent H$_2$ emission northwest of IRS 1. In this picture, high-mass star formation is progressing toward the southeast, 
and IRS 9 belongs to a second wave of star formation as it is placed further along the direction of propagation of the shock front.  These 
objects then represent two snapshots of the evolution of a massive young protostellar object in a particular cloud and are useful in 
addressing temporal evolution questions.

This paper is organized as follows.  In Section 2  we give details concerning how we obtained and reduced the data, followed by 
a detailed description in Section 3 of the molecular species detected.  The data analysis is described in Section 4 with emphasis on spectral modeling, and we give some interpretation to the results in Section 5.  We summarize 
these results and draw some conclusions in Section 6.

\section{Observations and Reductions\label{sec:obs}}

We observed NGC 7538 IRS 9 with the Texas Echelon Cross (X) Echelle Spectrograph (TEXES; \citealt{TEXES}) on the Gemini North 8 m 
telescope in 2007 October.  Similar observations of IRS 1 were carried out with TEXES at the NASA Infrared Telescope Facility (IRTF) 3 m 
telescope in 2001 June, 2001 November, 2002 September, 2002 December and 2005 November.  Circumstances for all IRS 9 observations 
are given in Table~\ref{irs9obstable}.  IRTF observations of IRS 1 form the basis for the work of \citet{Claudia}.  Additional observations of 
IRS 1 were made at Gemini in 2007 October to supplement previous TEXES data taken at the IRTF.
\begin{deluxetable*}{cccc}
\tabletypesize{\footnotesize}
\tablecolumns{4}
\tablewidth{0pt}
\tablecaption{Circumstances of TEXES observations of NGC 7538 IRS 9}
\tablehead{
& & Wavenumber & \\
\colhead{UT Date} & \colhead{Telescope} & \colhead{Center (cm$^{-1}$)} & \colhead{Included Features}
}
\startdata
2001 June 15 & IRTF & 768 & C$_2$H$_2$ ${\nu}_{5}$ $R$(15,16) \\
		        &          & 780 & [\ion{Ne}{2}] ($^{2}P_{1/2}$ $\to$ $^{2}P_{3/2}$), C$_2$H$_2$ ${\nu}_{5}$ \\
		        & 	  &         & $R$(21,22), HCN ${\nu}_{2}$ $R$(22,23) \\
2001 June 18 & IRTF & 744 & C$_2$H$_2$ ${\nu}_{5}$ $R$(5,6), HCN ${\nu}_{2}$ $R$(10)  \\
2001 June 26 & IRTF & 761 & C$_2$H$_2$ ${\nu}_{5}$ $R$(12,13), HCN ${\nu}_{2}$ $R$(15,16), \\
		       &	&	& HD, H$^{13}$CN ${\nu}_{2}$ $R$(18) \\ 
2001 June 28-29 & IRTF & 734 & C$_2$H$_2$ ${\nu}_{5}$ $R$(1), HCN ${\nu}_{2}$ $R$(6,7) \\
2007 October 21 & Gemini & 828 & HNCO ${\nu}_4$ $P$-branch, $^{14}$NH$_{3}$ ${\nu}_2$ s$P$(7,K) \\
			& 	& 1308 & $^{12}$CH$_{4}$ ${\nu}_4$ $R$(0), $^{13}$CH$_{4}$ ${\nu}_4$ $R$(1) \\
2007 October 25 & Gemini & 930 & $^{14}$NH$_3$ ${\nu}_2$ a$Q$(J,K) \\ 
			& 	& 743 & C$_2$H$_2$ ${\nu}_{5}$ $R$(5,6) \\
2007 October 26 & Gemini & 853 & $^{14}$NH$_3$ ${\nu}_2$ a$P$(4,K) \\
 			& 	& 1323 & $^{12}$CH$_{4}$ ${\nu}_4$ $R$(2), $^{13}$CH$_{4}$ ${\nu}_4$ $R$(4), \\
 			& 	& & C$_2$H$_2$ ${\nu}_{4}+{\nu}_{5}$ $P$(1-3)  \\
2007 October 27 & Gemini & 2055 & OCS ${\nu}_2$ $P$(12-22), $^{12}$CO $v$=1-0 $P$(21,22), \\
			&	&	& $^{13}$CO $P$(11,12), C$^{18}$O $P$(10,11), C$^{17}$O $P$(16)  \\
2007 October 28 & Gemini & 730 & C$_2$H$_2$ ${\nu}_{5}$ $Q$-branch \\
2007 October 30 & Gemini & 2085 & $^{12}$CO $v$=1-0 $P$(14-15), $^{13}$CO $P$(3,4), \\
			&	&	&	C$^{18}$O $P$(2,3), C$^{17}$O $P$(8,9) \\
\enddata
\label{irs9obstable}
\end{deluxetable*}

TEXES is a cross-dispersed grating spectrograph designed to operate at high spectral resolution (R = 75,000-100,000) over a range of mid-infrared wavelengths from 5 to 25 $\mu$m.  Light diffracts off a coarsely-ruled ``echelon'' grating \citep{echelon} at a large incidence angle 
with respect to normal at order $n$ $\sim$ 1500.  The light then diffracts off a cross-dispersion grating and is imaged onto a 256 $\times$ 
256 pixel Si:As detector as a series of orders.  At a given spectral setting between five and ten cross-dispersed orders are imaged onto the 
detector for a spectral coverage of about 0.5\%.  The continuity of orders is wavelength-dependent; below 11 $\mu$m order overlap is 
sufficient to yield continuous coverage over an entire spectral setting.  At wavelengths longer than 11 $\mu$m there are gaps between 
adjacent orders.

On Gemini the pixel scale of TEXES was $\sim$0$\farcs$14 while at the IRTF it was $\sim$0$\farcs$36.  The pixel scales also compare 
favorably with the diffraction limit at 10 $\mu$m of each telescope, 0$\farcs$70  and 0$\farcs$26 at the IRTF and Gemini, respectively.  In 
both cases the seeing disc is, at worst, critically sampled under good seeing conditions.  All observations were made at a resolving power 
$R$ ${\simeq}$ 80,000, or ${\Delta}v\sim$ 3-4 km s$^{-1}$ and the spectral coverage at each spectral setting was ${\Delta}{\lambda}\approx
$ 0.06 $\mu$m.  The spectral sampling of the TEXES array is about 1.0 km s$^{-1}$.   Slit widths were 0$\farcs$5 at Gemini and 1$\farcs$4 
at the IRTF.  The slit length was $\sim$4$\arcsec$ on Gemini and $\sim$10$\arcsec$ on the IRTF, varying with wavelength.   We alternated 
the telescope pointing between two positions on the sky in the slit direction, separated by $\sim$ 1-5$^{''}$, during an exposure sequence.  
Subtraction of adjacent ``nod'' positions removes the contribution to the signal from night sky emission.  The separation of points in the nod, 
or ``nod throw'', is determined by the slit length.

To correct for telluric atmospheric absorption we observed comparison objects whose spectra are close to blackbody over the wavelength 
range of observation.  At wavelengths beyond $\sim$ 8 $\mu$m, asteroids meet this requirement while providing the most flux of available 
object types.  Toward 5 $\mu$m, most asteroids provide insufficient flux to achieve a high signal-to-noise (S/N) ratio, so we used bright, hot 
Main Sequence stars such as Vega ($\alpha$ Lyr), Fomalhaut ($\alpha$ PsA) and Mirfak ($\alpha$ Per) as divisors.  We aimed to observe 
the divisor for a given target at nearly the same airmass as the target observation for better divisions.  Differences in airmass at the times of 
observations may be corrected after data reduction by taking the divisor spectrum to the power of the ratio of the airmasses.  Residual 
structure in divided spectra was removed by fitting and subtracting low-order polynomials.

Finally, given the high spatial resolution of the Gemini observations, we considered the possibility that TEXES might resolve the emitting/
absorbing region around IRS 9 if it were spatially extended.  We obtained a series of spatial-spectral scan maps of IRS 9 that allow us to 
place an upper limit on the linear size of any resolved structure.  The approach and results are discussed in detail in Section~\ref
{subsec:scanmap}.

\subsection{Data Reduction}

Our data reduction pipeline is described in detail in \citet{TEXES}.  Flux calibration is based on the approach of calibrating millimeter and 
submillimeter data in the manner of \citet{UH76}.  A typical sky calibration cycle involves collecting data on the sky and object with two 
positions of a rotating chopper blade mounted just above the entrance window to the TEXES Dewar: one painted flat black and one low-emissivity, polished aluminum (``shiny'') surface.Ê The result of dividing by the black-sky difference removes atmospheric absorption and 
serves as both a spatial and spectral flat field image provided that the black position, sky, and telescope temperatures are approximately 
equal.  We quote the \citet{TEXES} expression for the object intensity as a function of frequency,
\begin{equation}
I_{\nu} (object) \approx S_{\nu}(object-sky)\frac{B_{\nu}(T_{tel})}{S_{\nu}(black-sky)}
\end{equation}
where $S_{\nu}$ is the measured signal and $B_{\nu}$ is the blackbody function at the telescope temperature $T_{tel}$.  In the 
approximation that $I_{\nu}$(black)-$I_{\nu}$(sky) = $B_{\nu}$($T_{tel}$)$e^{-{\tau}_{sky}}$, this procedure corrects for sky absorption.  
Wavelength calibration is achieved by using the known wavelengths of night sky emission lines; expressed in velocity units, this method is 
accurate to $\sim$1 km s$^{-1}$.
 
Pipeline software written in FORTRAN performs the reduction steps resulting in a wavelength- and flux-calibrated object spectrum along with 
an estimate of the noise, the sky transmission spectrum, and the 2-D unextracted slit image.  The program also corrects for optical distortion, 
removes cosmic ray spikes and fixes bad pixels.  The systematic accuracy of radiometric calibrations is $\sim$25\%, ignoring telescope 
pointing uncertainties.

\section{Description of the Data\label{sec:data}}

We observed IRS 9 in the lines of seven molecules at the settings summarized in Table~\ref{irs9obstable}.  The selection of settings was 
motivated by the desire to include as many species and individual lines as possible given the severity of telluric atmospheric absorption at a 
particular setting.  In particular, variations in the H$_2$O column over Mauna Kea often influenced the settings observed on a particular 
night, as many of the settings are sensitive to the intensities of telluric water lines.  The systemic velocity of IRS 9 and motion of the Earth 
with respect to the Local Standard of Rest (LSR) were taken into account in choosing settings in which certain lines were shifted either into 
or away from nearby telluric absorption features, and in many cases required trade-offs in order to maximize efficiency and scientific return.  
Details of the observations for specific molecules follow.

\subsection{C$_2$H$_2$}

\begin{figure}
\leavevmode
\includegraphics[width=0.48\textwidth]{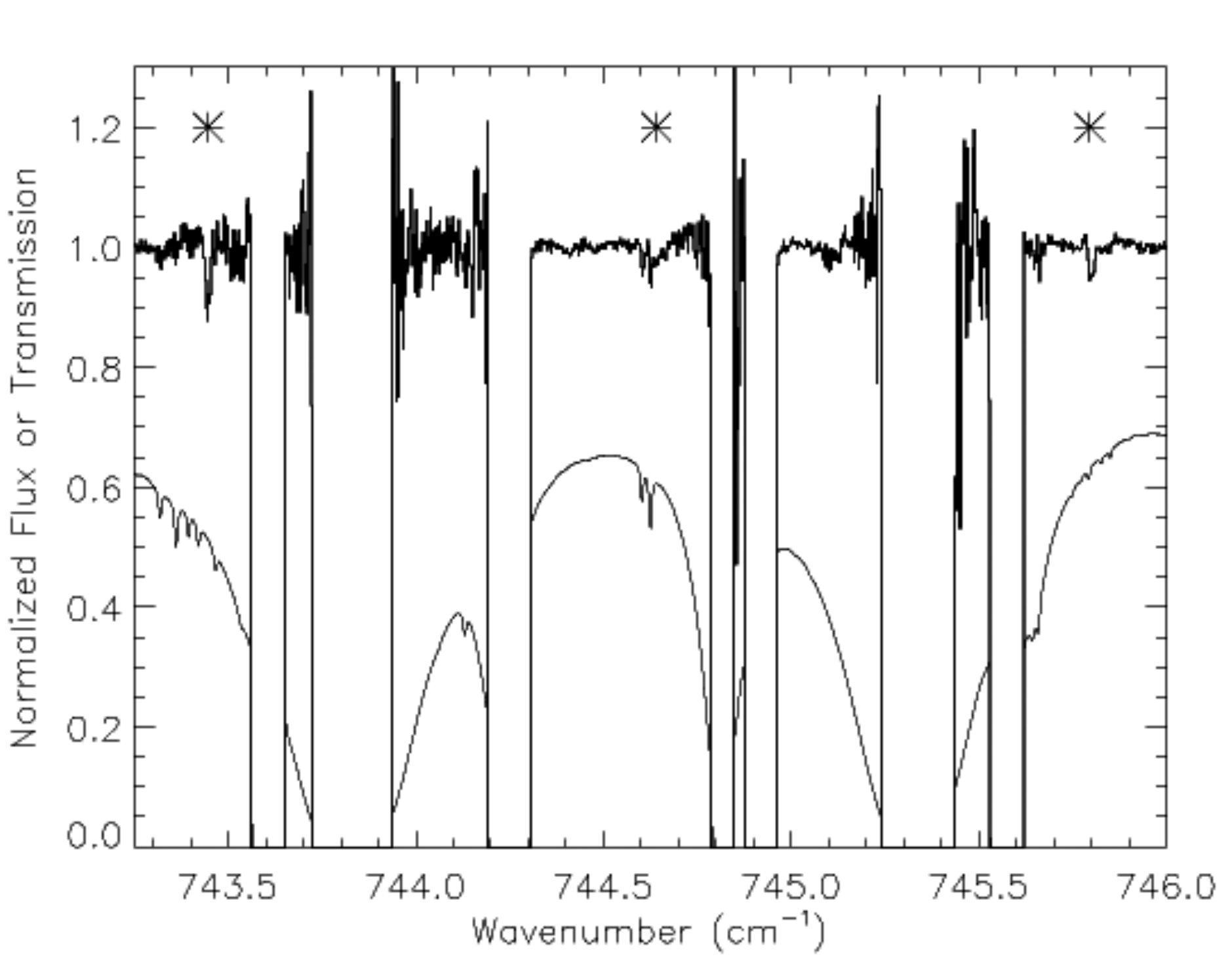}
\caption{TEXES spectrum toward NGC 7538 IRS 9 showing features due to C$_2$H$_2$ and HCN.  From left to right, the asterisks mark the 
positions of C$_2$H$_2$ ${\nu}_5$ $R$(5), HCN ${\nu}_2$ $R$(10), and C$_2$H$_2$ ${\nu}_5$ $R$(6).  The upper trace shows the data, 
while the lower curve indicates the relative atmospheric transmission on the same scale.  The HCN ${\nu}_2$ $R$(10) line is a detection 
despite the telluric lines at that wavenumber not completely dividing out.  Gaps in the spectrum occur between echelon grating orders and 
where the telluric absorption is too great to be divided out.}
\label{acet_rbranch}
\end{figure}

\begin{figure*}\centering
\leavevmode
\includegraphics[width=0.8\textwidth]{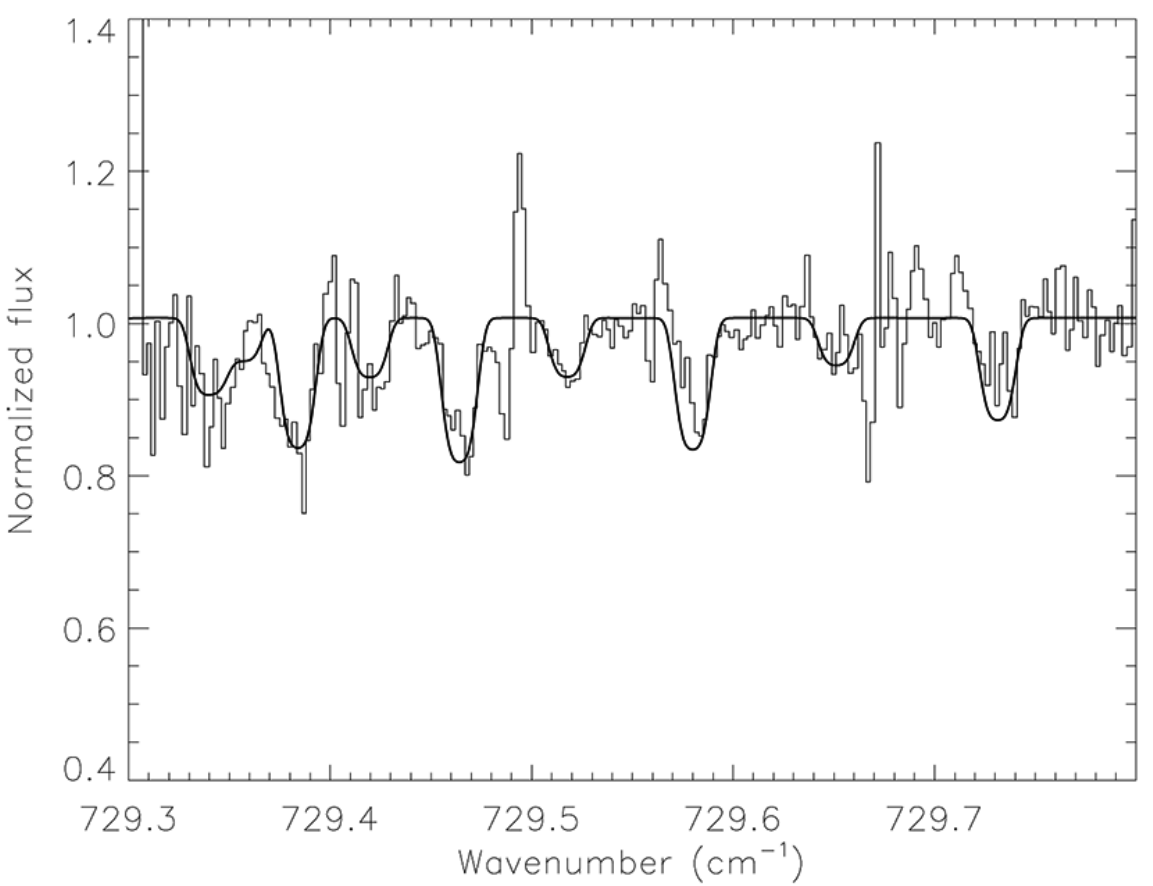}
\caption{Spectrum of the ${\nu}_5$ $Q$ branch of C$_2$H$_2$ observed with TEXES toward NGC 7538 IRS 9 (thin lines).  A fit is 
superimposed from an LTE pure absorption model with $T$ = 100 K, $b$ = 1 km s$^{-1}$, and $N$(C$_2$H$_2$) = 2 $\times$ 10$^{15}$ 
cm$^{-2}$ (thick line).  The synthetic spectrum has been convolved with the TEXES instrumental line shape function.}
\label{acetsampspec}
\end{figure*}

At Gemini, two lines in the acetylene ${\nu}_5$ symmetric bending mode $R$ branch were observed ($J$=5,6) along with approximately ten 
lines in the ${\nu}_5$ $Q$ branch.  The $R$-branch lines are shown in Figure~\ref{acet_rbranch}, while Figure~\ref{acetsampspec} shows 
the $Q$-branch spectrum along with a simple LTE fit at a single rotational temperature of 100 K.  This is discussed further in Section 4.  Each acetylene line appears to consist of a single component.  The $R$(5) line is stronger than $R$(6), as is expected due 
to its three-times-greater nuclear statistical weight.  $R$(6) appears to have a weak redshifted emission component at an LSR velocity of -54 
km s$^{-1}$ although this effect may be due to a poor fit to the continuum there.  This component can be more easily seen in the top-center
panel of Figure~\ref{fits_p1}.  Earlier IRTF observations of the ${\nu}_5$ $R$(1) line (wavenumber center 734 cm$^{-1}$) in 2001 June 
revealed a positive detection, but more importantly, higher $J$ settings $R$(15,16) (768 cm$^{-1}$) and $R$(21,22) (780 cm$^{-1}$) were 
non-detections.  The latter setting included the $^{2}P_{1/2}$ $\to$ $^{2}P_{3/2}$ fine structure line of [\ion{Ne}{2}]; consistent with the weak 
free-free emission observed toward IRS 9, this line was also a non-detection and is discussed in detail in Section~\ref{subsec:neflux}.  
Finally, although the positions of lines of the $^{13}$C$^{12}$CH$_2$ isotopologue fell in some of our spectral settings, we did not detect 
any.

\begin{figure}
\leavevmode
\includegraphics[width=0.5\textwidth]{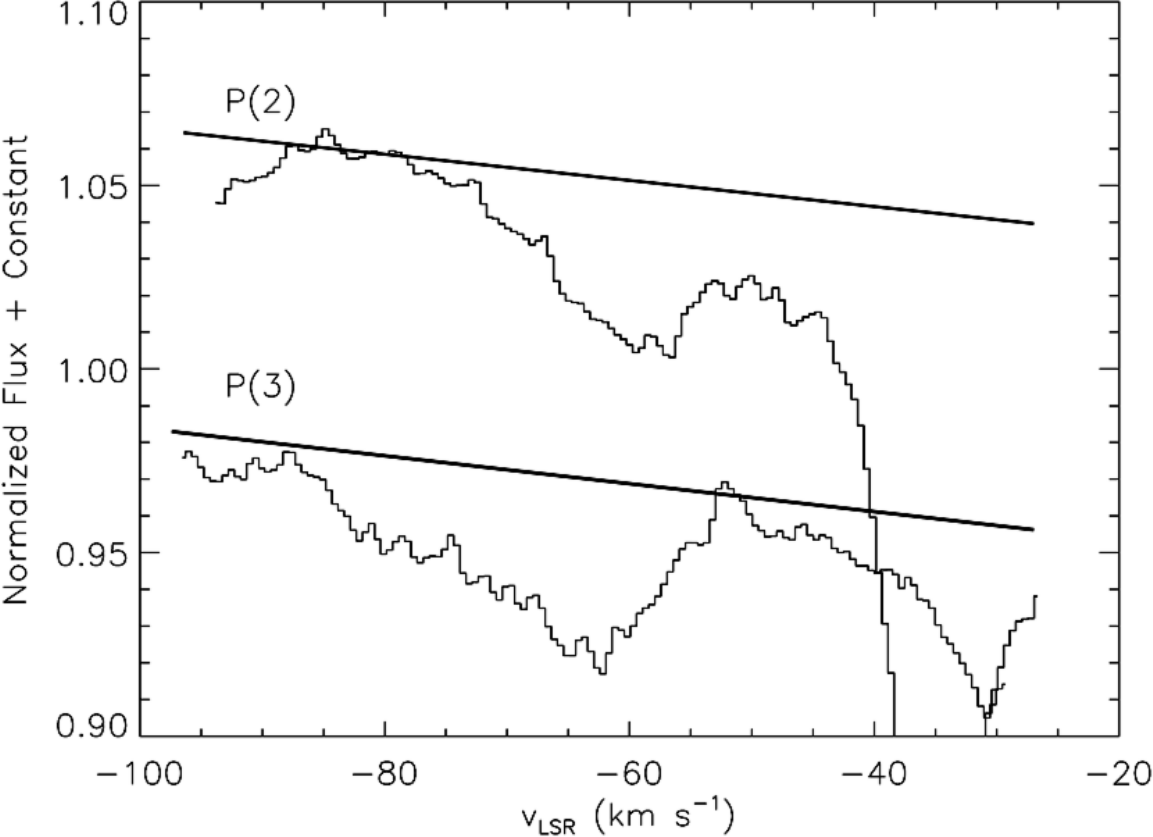}
\caption{Spectra of two $P$-branch lines of the ${\nu}_{4}+{\nu}_{5}$ combination mode of C$_2$H$_2$ shown on an LSR velocity scale. 
The two spectra have been offset vertically for clarity, and the continuua are indicated with heavy straight lines.  Both spectra are affected by 
strong, poorly-corrected telluric absorption at small blueshifts.}
\label{combination_spec}
\end{figure}

We also did not detect any lines in the 2${\nu}_5-{\nu}_5$ ``hot bands'' of acetylene toward IRS 9.  However, the hot-band lines in the 
spectral settings observed were not particularly favorable for detection, so we cannot draw any strong conclusions about the ${\nu}_5$ 
population from these observations.  The observation of ${\nu}_{4}+{\nu}_{5}$  combination mode lines, shown in Figure~\ref
{combination_spec}, is of particular interest.  This is because it allows a measurement of the C$_2$H$_2$ abundance with intrinsically 
weaker lines at different wavelengths from the ${\nu}_5$ band lines.  Since the ${\nu}_4+{\nu}_5$ line strengths are $\sim$5 times weaker 
than the ${\nu}_5$ line strengths, absorption in the ${\nu}_4+{\nu}_5$ lines may be dominated by high column density clumps of gas.
\begin{figure}\centering
\leavevmode
\includegraphics[width=0.5\textwidth]{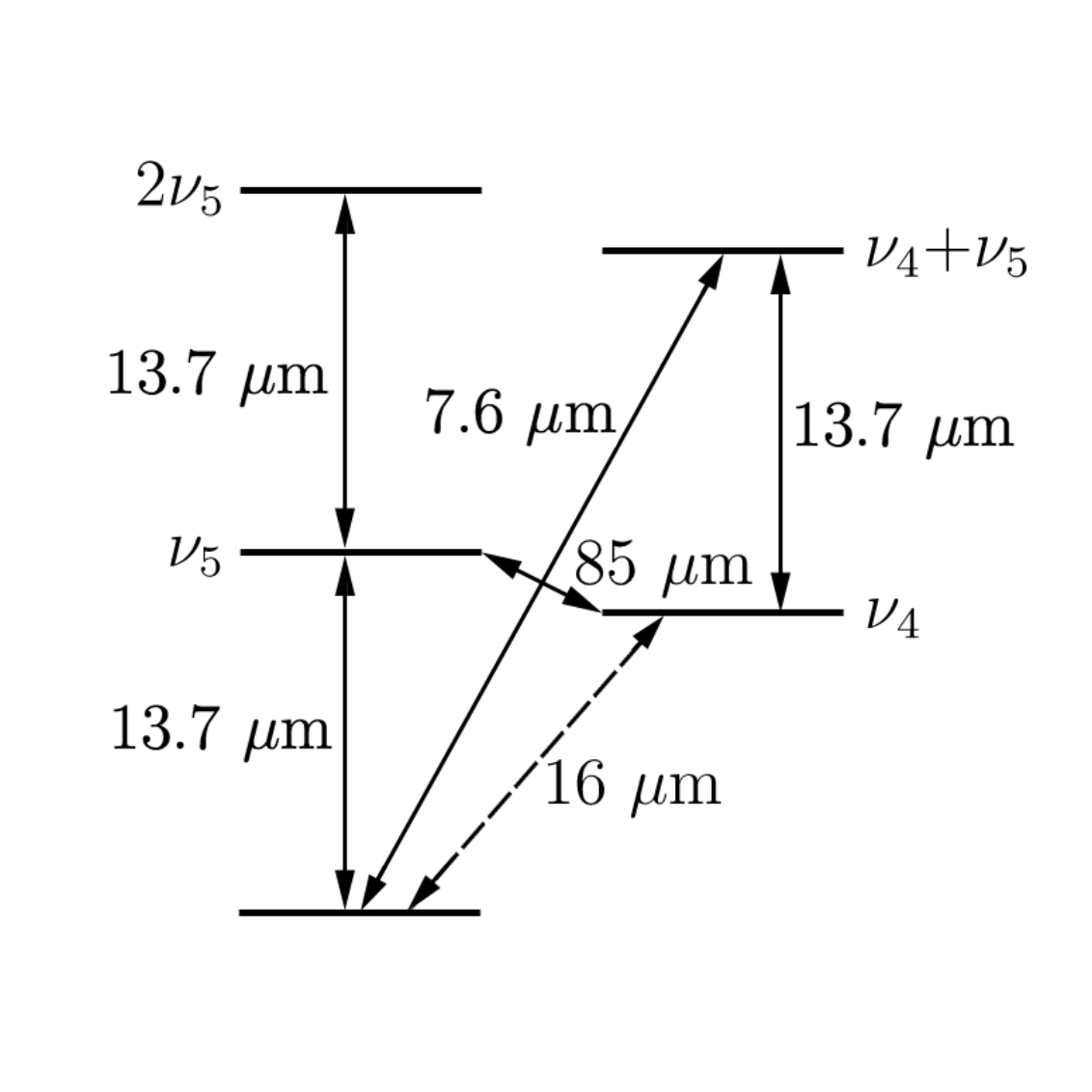}
\caption{Energy level diagram for some mid-infrared-active, ro-vibrational transitions of C$_2$H$_2$.  Radiatively-allowed transitions are 
denoted with solid lines and  collisionally-allowed-only transitions with dashed lines.  The excited vibrational levels are split by interactions 
between the rotational angular momentum and the angular momentum of the bending-mode vibration; however, the energy differences 
between these levels are too small to be seen at the resolution of the diagram.}
\label{c2h2_levels}
\end{figure}  

An energy level diagram indicating relevant vibrational levels of C$_2$H$_2$ is shown in Figure~\ref{c2h2_levels}.  The ${\nu}_5$ mode is 
the symmetric bending mode in which the H atoms move together.  In the ${\nu}_{4}$ antisymmetric bending mode, they move in opposition, 
resulting in no dipole moment. Therefore, transitions that change the ${\nu}_5$ quantum number are relatively strong, with $A_{ul} \sim 
10$ s$^{-1}$.  These include the ${\nu}_5$, 2${\nu}_5-{\nu}_5$ (between 2${\nu}_5$  and ${\nu}_5$ ), and ${\nu}_{4}+{\nu}_{5}-{\nu}_{4}$ 
(between ${\nu}_{4}+{\nu}_{5}$ and ${\nu}_{4}$) transitions.  Transitions that only change ${\nu}_{4}$ do not occur radiatively, but transitions 
changing both ${\nu}_{4}$ and ${\nu}_5$, including ${\nu}_{4}+{\nu}_{5}$ and ${\nu}_{5}-{\nu}_{4}$, do occur although they are $\sim10$ 
times weaker than those which only change ${\nu}_5$.  Because of these selection rules, 7.6 $\mu$m photons can be absorbed in ${\nu}_
{4}+{\nu}_{5}$ lines, but molecules excited into the ${\nu}_{4}+{\nu}_{5}$ level most often decay to the ${\nu}_{4}$ level from which they can 
only decay to the ground vibrational level collisionally.

The ratio of the ${\nu}_4+{\nu}_5$ $P$(3) and $P$(2) lines is difficult to determine due to poor telluric division, but appears to be less than 
the 9:2 ratio expected from their rotational and nuclear statistical weights.  This probably indicates saturation of the lines in spite of their 
intrinsic weakness relative to the ${\nu}_5$ transitions.  We also note that the spectral region containing the ${\nu}_{4}+{\nu}_{5}$ lines is 
contaminated by dense forests of HNO$_3$ and SO$_2$ lines, giving the superficial appearance of bad flat-fielding.  Another difference 
between the ${\nu}_5$ and the ${\nu}_{4}+{\nu}_{5}$ transitions involves the shapes of the lines;  the ${\nu}_{4}+{\nu}_{5}$ lines are 
more asymmetric, with blue wings extending to nearly -90 km s$^{-1}$.  In this regard they resemble the P Cygni-like shapes of the CO lines 
we observed toward IRS 9 (see Section~\ref{sec:co}).  The ${\nu}_{4}+{\nu}_{5}$ lines may result from absorption in a high column density 
component with a small covering factor that is strongly saturated.  Or they may be formed in highly dust-obscured gas that is more visible at 
7.6 than 13 $\mu$m due to the lower dust opacity off of the silicate absorption feature.  The presence of the similar -80 km~s$^{-1}$ 
absorption feature in the CO lines supports this possibility.  Alternately, the ${\nu}_{4}+{\nu}_{5}$ line strengths relative to the ${\nu}_5$ lines 
may be affected by other radiative transfer effects, as we discuss  in Section~\ref{subsec:nh3scheme}.

\subsection{HCN}

Three of our spectral settings contain $R$-branch lines of the ${\nu}_2$ bending mode of HCN with centers at 734 cm$^{-1}$ (6 $\le$ $J$ $
\le$ 7), 745 cm$^{-1}$ ($J$=10), and 761 cm$^{-1}$ ($J$=16).  The $R$(6) line fell into a region where atmospheric transmission is not 
favorable, and $R$(7) appears to be a marginal detection because of a weaker, nearby atmospheric feature.  $R$(10) (Figure~\ref
{acet_rbranch}) is a probable detection.  $R$(16) falls close to the $R$(13) line of C$_2$H$_2$ but is sufficiently separated from the 
acetylene line to prevent blending and is the only clean HCN line detected.   The $R$(15) line of HCN fell just outside the first order of the 
761 cm$^{-1}$ setting.  The S/N of the HCN observations is generally lower than that of C$_2$H$_2$ ${\nu}_5$ $R$ branch, but the HCN 
lines are clearly the broader of the two.  The mean full width at half maximum (FWHM) of the four C$_2$H$_2$ $R$-branch lines is
6.5$\pm$1 km s$^{-1}$, whereas the HCN $R$(16) line FWHM is $\sim$15 km s$^{-1}$.  Finally, lines of the H$^{13}$CN isotopologue are 
also found in these settings but not detected: $R$(9) in 734 cm$^{-1}$  and $R$(18) in 761 cm$^{-1}$.  $R$(18) falls into a gap between 
orders and $R$(9) was affected by poor atmospheric transmission.

\subsection{CH$_4$}

\begin{figure}
\leavevmode
\includegraphics[width=0.5\textwidth]{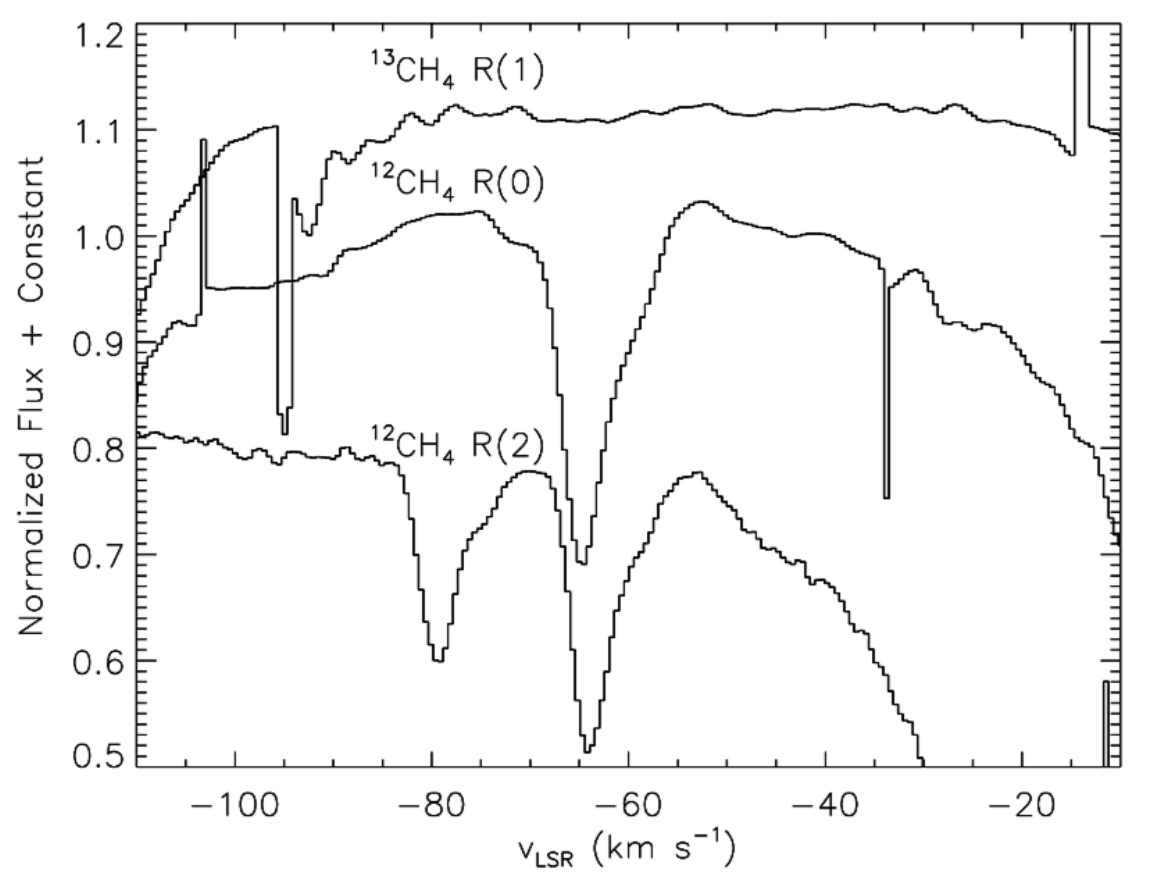}
\caption{Spectra of three features in the ${\nu}_4$ band of CH$_4$ toward IRS 9, shown on an LSR velocity scale.  For the spectrum 
containing the $R$(2) lines, the velocity scale has been set to the LSR value for the stronger component (rest frequency 1322.085 cm$^{-1}$).  The individual spectra have been offset vertically for clarity.}
\label{ch4sampspec}
\end{figure}

We obtained data on three ${\nu}_{4}$ $R$-branch lines of CH$_4$: the $R$(0) line at 1311.43 cm$^{-1}$ and the two $R$(2) lines at 
1322.083 cm$^{-1}$ and 1322.152 cm$^{-1}$.  In addition, the settings for these lines contained the positions of the $R$(1) and $R$(4) lines 
of $^{13}$CH$_4$, but both of these lines were clear non-detections.   The $^{12}$CH$_4$ lines show clear asymmetries indicative of 
multiple velocity components.  Each feature has a principal component corresponding to greatest absorption at $v_{LSR}$ = -65.0 km s$^
{-1}$ with additional components on the red wing of this feature separated in velocity by up to 6.5 km s$^{-1}$.  The spectra are plotted in 
Figure~\ref{ch4sampspec}.

\subsection{NH$_3$}

\begin{figure}\centering
\leavevmode
\includegraphics[width=0.5\textwidth]{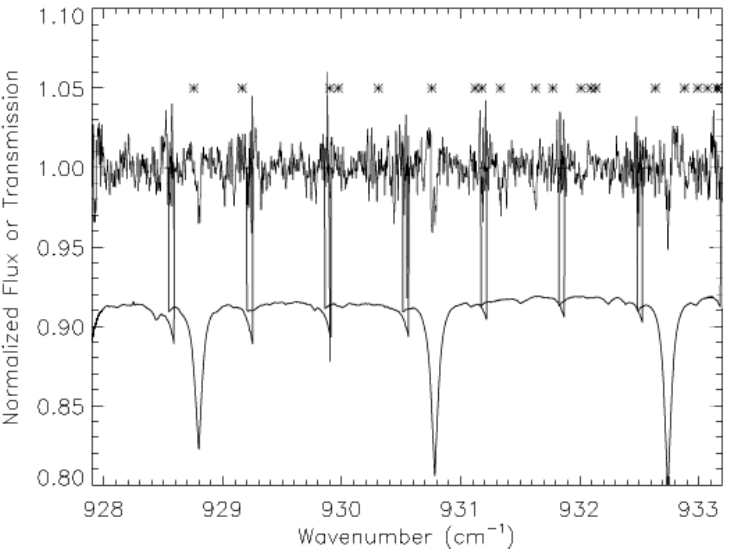}
\includegraphics[width=0.5\textwidth]{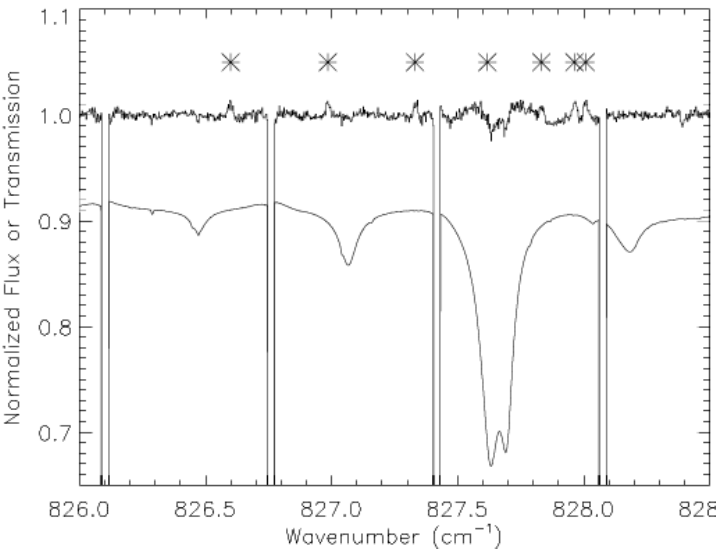}
\caption{TEXES spectra toward NGC 7538 IRS 9 at the position of the $^{14}$NH$_3$ ${\nu}_2$ antisymmetric $Q$ branch (top panel) and 
symmetric $P$(7,$K$) branch (bottom panel).  The upper curve in each plot shows the data, while the lower curve in each shows the relative 
atmospheric transmission on the same scale.  Asterisks mark positions of the NH$_3$ lines.  Both spectra have been corrected for the 
Earth's motion relative to the LSR, and the position of the NH$_3$ line markers shifted to the systemic LSR velocity of approximately -60 km 
s$^{-1}$.}
\label{nh3branches}
\end{figure}

Two spectral settings observed contain a total of ten $P$-branch lines of ammonia, three in the ${\nu}_2$ ``umbrella'' mode antisymmetric $P$(4,$K$) group and seven in the ${\nu}_2$ symmetric $P$(7,$K$) group.  Of the possible lines in these settings, only a$P$(4,2) was not 
observed because of its proximity to a strong telluric atmospheric feature.  Almost all of the NH$_3$ lines share a characteristic P Cygni-like 
profile, with an emission peak near the systemic velocity, around -60 km s$^{-1}$, and weak blueshifted absorption separated in velocity by 
2.5 to 10 km s$^{-1}$.  This may be compared to the -56.5 and -59.8 km s$^{-1}$  absorption components seen in mid-infrared NH$_3$ 
absorption transitions toward IRS 1 \citep{Claudia} and a value of -60.0 km s$^{-1}$ for the (3,3) rotational transition of NH$_3$ in Very 
Large Array (VLA) observations of IRS 1 at centimeter wavelengths \citep{HWJ84}.  The degree of absorption toward IRS 9 relative to 
emission varies as well; the highest $K$ lines in the s$P$(7,$K$) series show no absorption at all despite the fair S/N of the data.  Finally, 
we observed a setting near 930 cm$^{-1}$ containing the positions of the ${\nu}_2$ a$Q$-branch lines (1 $\leq$ $J$ $\leq$ 9); the data 
show weak absorption features at some of the indicated points in Figure~\ref{nh3branches}.  For comparison, a similar plot of the s$P$(7,$K
$) branch spectrum is also shown in the figure.  We did not observe lines in the ${\nu}_2$ $R$-branch of NH$_3$ because of their proximity 
to regions of significant telluric atmospheric absorption.  Also, the $R$-branch lines are located in a spectral region of rapidly increasing 
absorption due to the 9.7 $\mu$m silicate dust feature.  This fact significantly impacts our interpretation of the NH$_3$ $P$- and $R$-branch 
lines, as we discuss in Section~\ref{subsec:nh3scheme}.

\subsection{CO\label{sec:co}}

\begin{figure}
\centering
\includegraphics[scale=0.75]{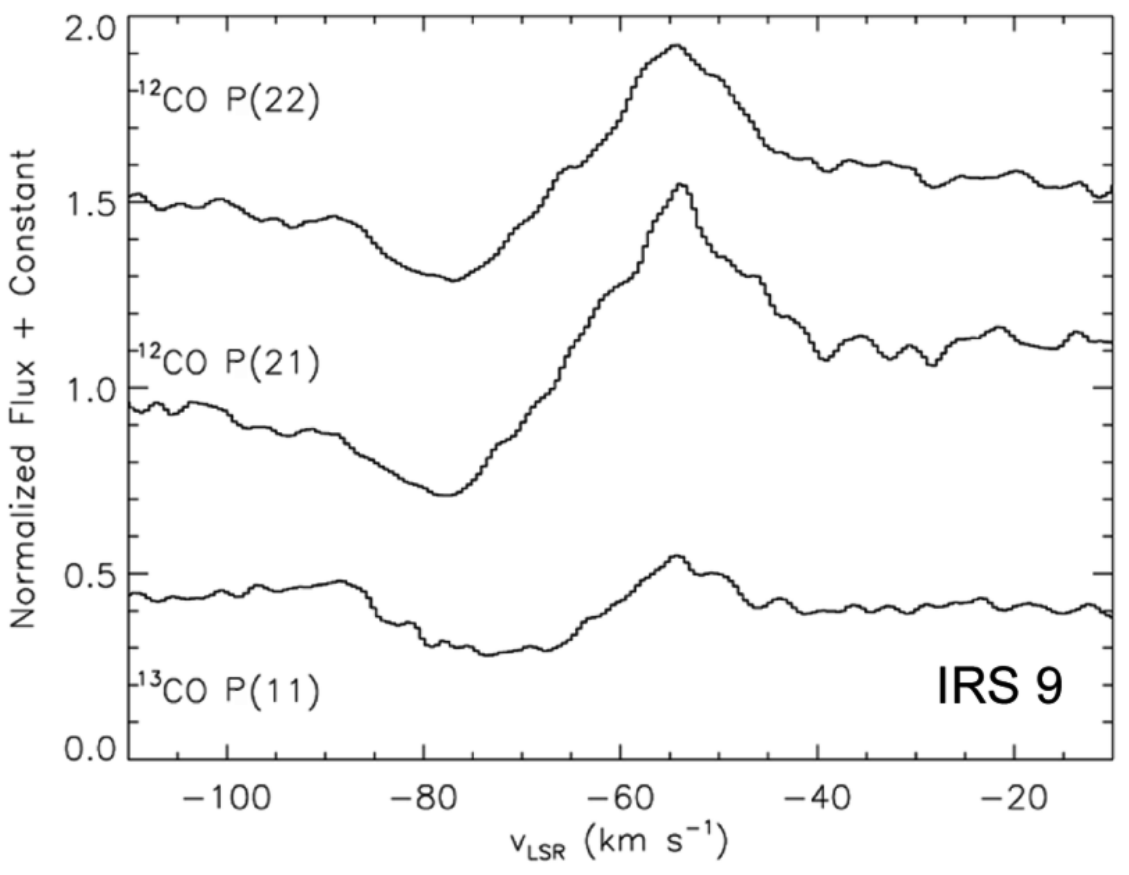} \\
\includegraphics[scale=0.75]{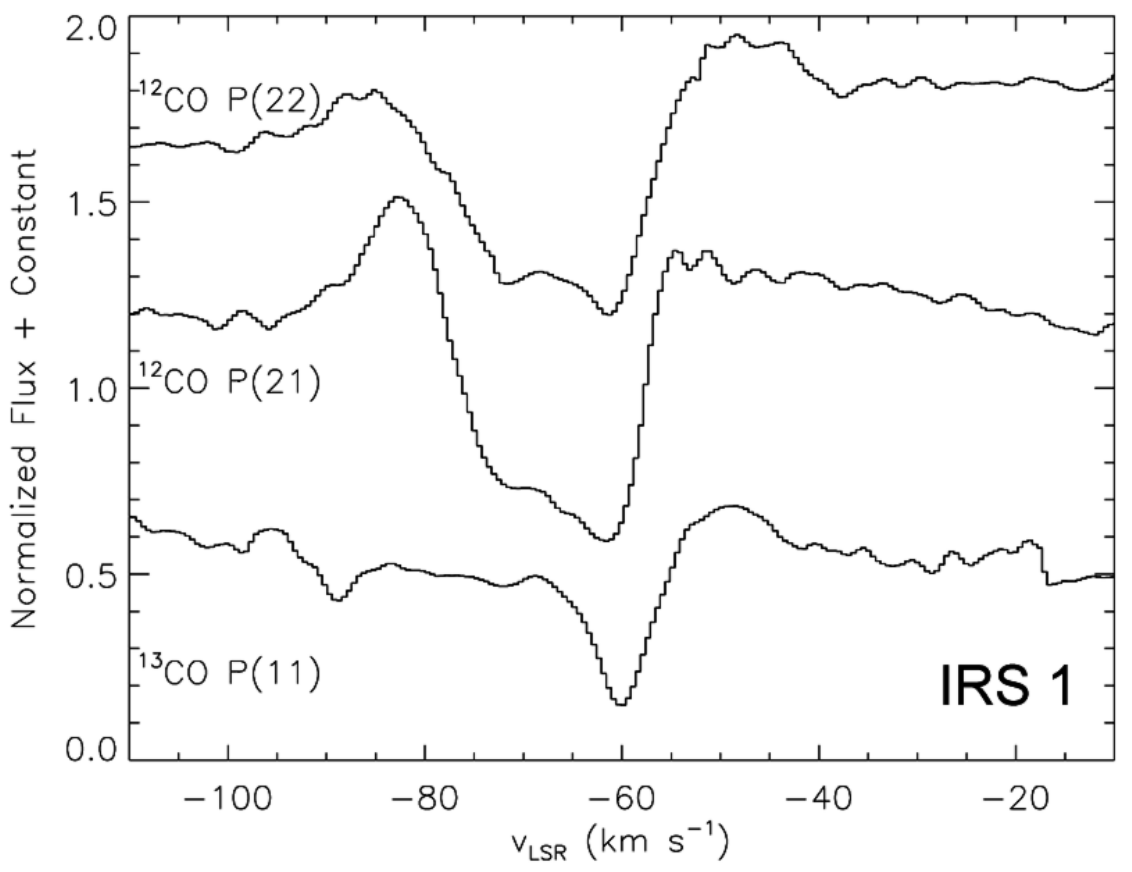}
\caption{The upper panel shows spectra of two $P$-branch lines of $^{12}$CO and one $P$-branch line of $^{13}$CO toward IRS 9 
obtained with TEXES.  The lower panel shows the same spectral settings observed toward IRS 1.  In both cases the motion of the Earth with 
respect to the LSR has been removed from the velocity scales.}
\label{irs91co}
\end{figure}

We observed two settings in the 5 $\mu$m $v$ = 1-0 band of CO, which included a number of $P$-branch lines of several isotopologues.  
These settings were centered approximately at 2055 cm$^{-1}$ and 2085 cm$^{-1}$.  Strongest among the detected features were the $^
{12}$CO lines with $J$ = 14, 15, 21 and 22.  Each line exhibited a strong P Cygni profile with emission and absorption components 
separated by $\sim$20 km s$^{-1}$.  Each component was broader than those of any line of any other molecular species we observed, with 
line widths of 10-20 km s$^{-1}$.  In each case the strength of the emission was approximately equal to the strength of the absorption.  The 
features are consistent with an expansion or outflow and have been previously interpreted as such by \citet{Mitchell91}.  Observations 
of IRS 1 were also made at these settings; we find a different profile altogether with a double-bottomed absorption feature in $^{12}$CO, 
weaker absorption on the blue wing and a slight emission bump on the red wing of each feature.  Spectra of both objects are shown in 
Figure~\ref{irs91co}.

Four $^{13}$CO $P$-branch lines with $J$ = 3, 4, 11, and 12 were also detected in the spectrum of IRS 9.  Of these, $P$(11) was a clear 
detection near the middle of a spect7ral order and away from contamination by telluric atmospheric lines, $P$(12) was noisy and found very 
near the edge of a spectral order, and $P$(8) and $P$(9) are clearly detected, but blended slightly with two nearby C$^{18}$O features.  
Additional $^{13}$CO lines, $P$(11) and $P$(12), are located in the spectral setting at 2055 cm$^{-1}$ but occur in regions of strong telluric 
atmospheric absorption and thus were not detected.  We also searched for lines attributable to C$^{17}$O but did not confidently detect any; 
this is not surprising given the low expected abundance of C$^{17}$O.  For reference, \citet{Smith09} quote $N$($^{12}$C$^{16}$O)/$N$($^
{12}$C$^{17}$O) of 2800$\pm$300 toward the T Tauri star VV CrA, obtained at a spectral resolution similar to our TEXES observations.  
They compare this to the Local ISM value of 2005$\pm$155 \citep{Wilson99} 

\begin{figure*}
\centering
\includegraphics[width=0.8\textwidth]{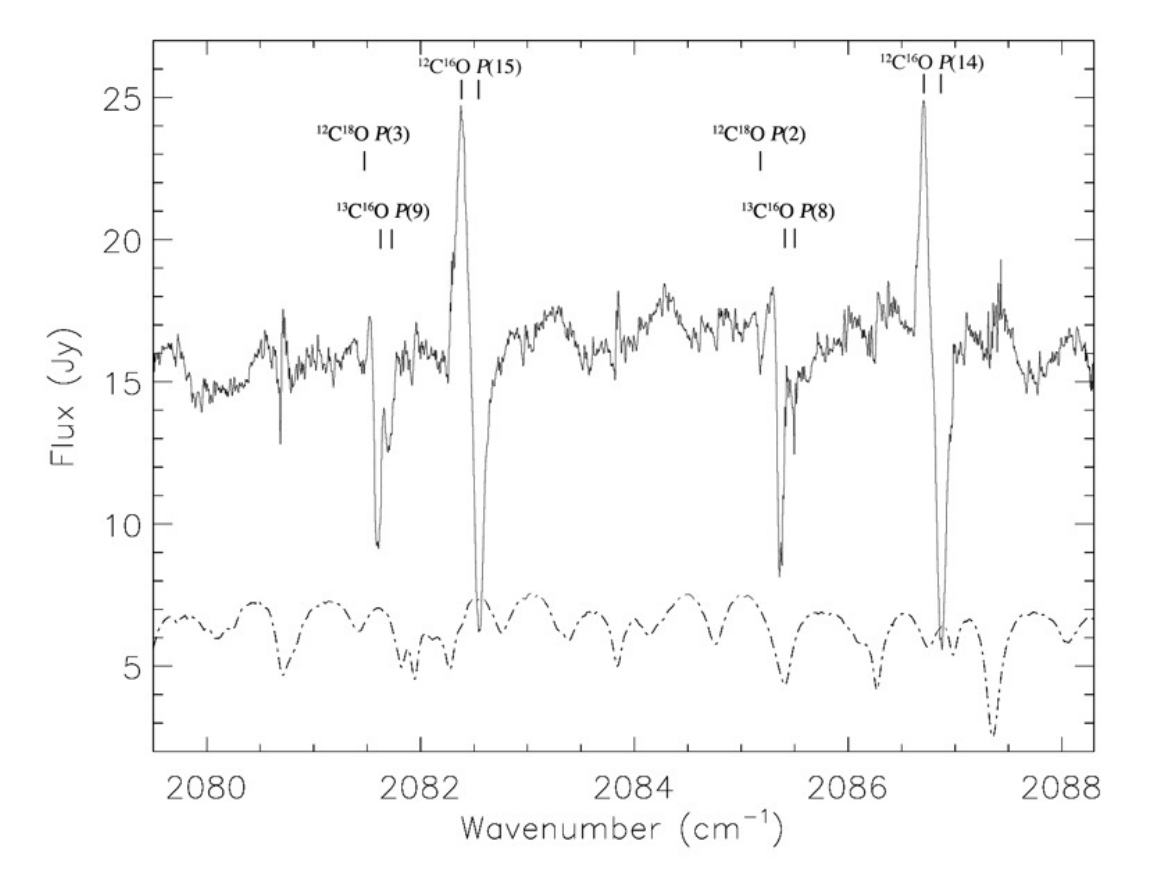}
\caption{TEXES spectrum toward NGC 7538 IRS 9 centered near the 5 $\mu$m rotational lines of several isotopologues of CO.  The data 
are shown in the upper trace (solid line) and the relative atmospheric transmission in the lower trace (dot-dashed line) multiplied by a factor 
of 10.  Some line identifications are shown.}
\label{co2085}
\end{figure*}

The CO lines seen toward both IRS 9 and IRS 1 have different line shapes in $^{12}$CO and $^{13}$CO.  In the case of IRS 1, the $^
{13}$CO features show only single components in the deepest absorption, while toward IRS 9 the $^{13}$CO $P$(11) line and $^{12}$CO 
lines show the pronounced P Cygni profiles.  The $^{13}$CO lines in the 2085 cm$^{-1}$ setting clearly show multiple components 
suggestive of P Cygni profiles in which emission is weak relative to absorption, and there are at least two distinct absorption units.  In turn, the 
strongest absorption components show a weak, double-bottomed structure like the $^{12}$CO lines in IRS 1.  The multiple absorption units 
may appear in the profiles of the $^{12}$CO features in the same spectral setting; weak ``shoulders'' on the blue wing of the absorption 
component are consistent with additional absorption suffering contamination by the primary components.  Such shoulders are 
also noticeable on the red wings of the emission features of $^{12}$CO .  These details can be seen in  Figure~\ref{co2085}.

The detection of $^{13}$CO toward IRS 9 with line depths comparable to those of $^{12}$CO strongly indicates saturation of both 
isotopologues.  Useful columns may however be obtained from rarer variants such as C$^{18}$O, but these lines are too weak to be 
measured accurately in the present spectra.  It would also be of interest measure the $^{12}$C/$^{13}$C ratio toward IRS 9 for comparison 
with the value of 45 toward IRS 1 reported by \citet{Claudia}.  As we did not observe the $^{13}$C isotopologue of C$_2$H$_2$ in our 
spectra of IRS 9, the CO observations would offer a way to measure the ratio if the lines were not saturated.

\subsection{OCS}

\begin{figure*}
\leavevmode
\begin{center}
\includegraphics[width=0.75\textwidth]{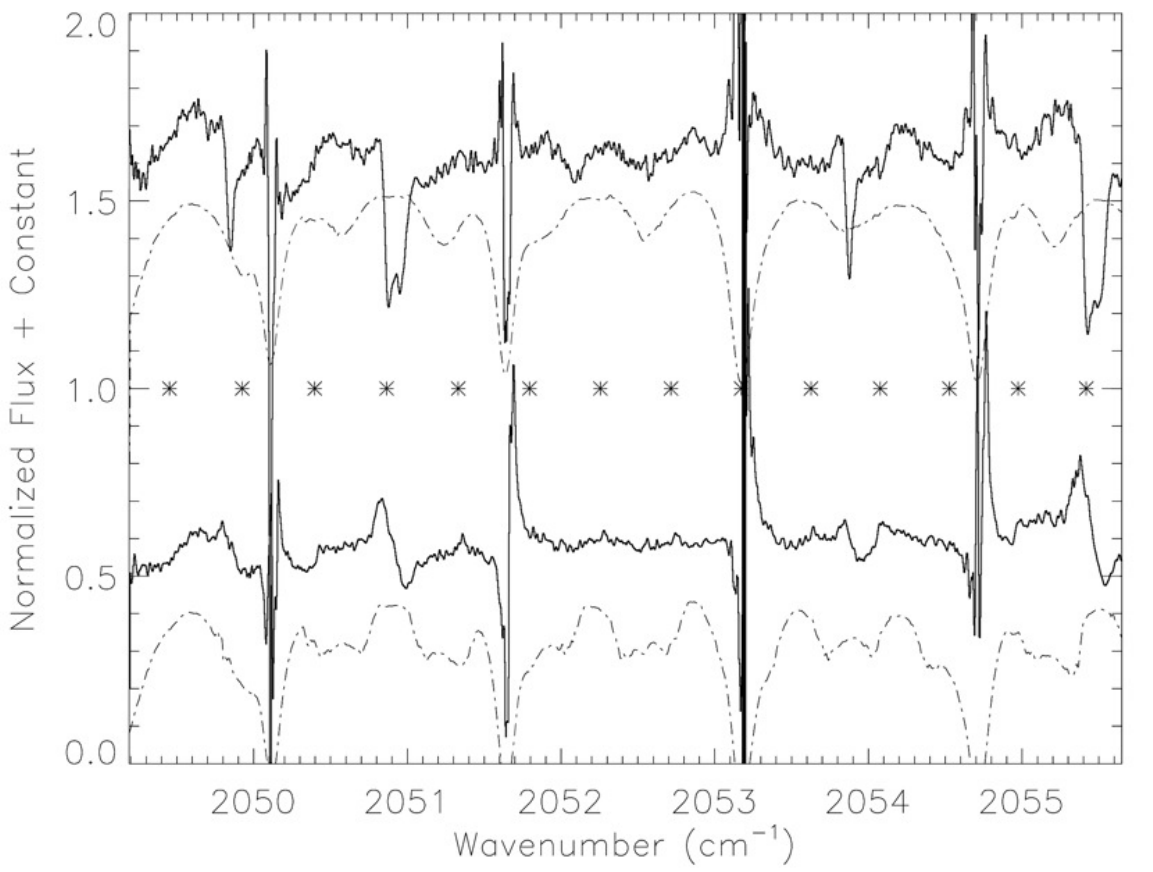}
\includegraphics[width=0.75\textwidth]{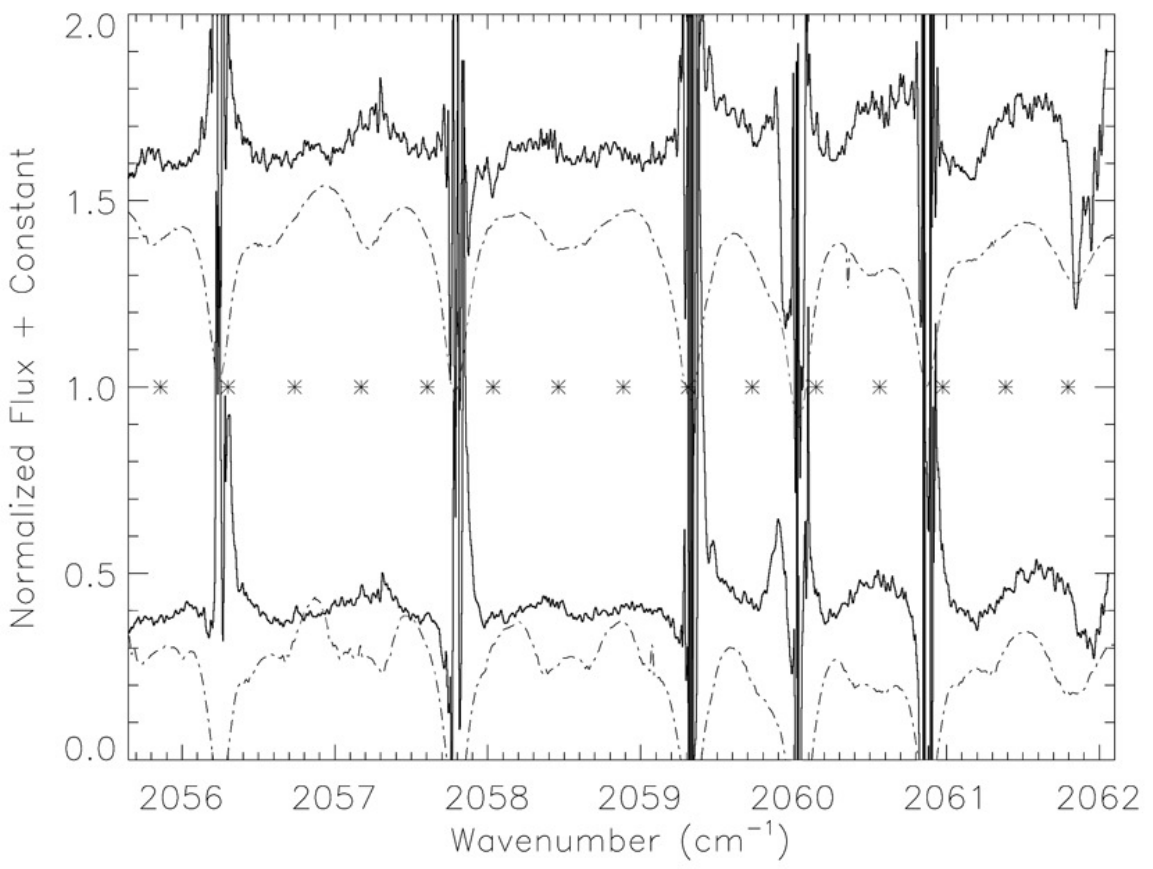}
\end{center}
\caption{The TEXES spectra of NGC 7538 IRS 1 (top trace in each panel) and IRS 9 (bottom trace) near features in the $P$-branches of $^
{12}$CO and $^{13}$CO; heavy solid lines are the object spectra and light dot-dash lines are the sky spectra offset for clarity.  The 
wavenumber scale has been corrected to a velocity reference at rest in each object's frame, and the line centers of OCS features, marked 
with asterisks, have been velocity shifted appropriately.}
\label{cospec}
\end{figure*}

The $^{12}$CO setting at 2055 cm$^{-1}$ we observed also contains the positions of a number of $P$-branch lines of the ${\nu}_2$ 
stretching mode of OCS (12 $\leq$ $J$ $\leq$ 22), but we cannot confidently claim detections of any of them toward IRS 9.  Periodic 
structure in the continuum of our 5 $\mu$m spectra may be broad OCS features, although it is more likely the structure results from flat-
fielding residuals.  For reference, we have overplotted the locations of the rest-frame OCS line centers over the data in Figure~\ref{cospec}.  
It is unclear whether we should have observed lines of OCS in the gas phase, given the strong detections of CO nearby.  Low temperatures 
may inhibit desorption of OCS from grains, but the theoretical grounds for predicting OCS abundances are questionable \citep{Doty04}.  The 
signature of an outflow in the spectra of CO hints at possible excitation through shocks.  We might expect that OCS would be similarly 
excited but there is no support for this in our data.  This is contrasted with the results of \citet{Evans91}, who detected gas-phase OCS toward 
Orion IRc2; for reasons given in Section 5, it is impossible to know whether this is due to temperature differences or real 
abundance variations between the IRS 9 and IRS 1.

\subsection{HNCO}

We looked for ${\nu}_4$ bending mode HNCO IL-0 $P$-branch lines toward IRS 9 in the TEXES spectral setting that included s(7,$K$) lines 
in the $P$ branch of ammonia.  None of the three lines included in this setting (2 $\leq$ $J$ $\leq$ 4) was detected, although there may be 
HNCO $Q$-branch absorption at 827.7 cm$^{-1}$, obscured by telluric interference.  Given the strong observed ice absorption at 4.62 $\mu
$m toward IRS 9 \citep{tielensxcn}, now commonly attributed to OCN$^{-}$, and a lack of gas-phase HNCO absorption, it seems most likely 
that temperatures in the outer envelope of IRS 9 are sufficiently low to prevent OCN$^{-}$ desorbing from grains in detectable quantities.  
Alternately, gas-phase OCN$^{-}$ might exist at H (or H$_2$) abundances insufficient to result in considerable HNCO production.

\section{Analysis and Modeling\label{sec:analysis}}

We fit synthetic spectra to the TEXES data to derive excitation parameters and column densities for the various molecular species observed.  
A two-step process was devised in which rough estimates of the parameters were made from ``by eye'' fits of synthetic spectra to our data; 
these values were then used as the initial guesses in our fitting program which iterates values of hundreds parameters to achieve the best fit 
to the data by minimizing ${\chi}^{2}$.  While our ${\chi}^{2}$ code is robust in finding solutions even when starting far from the global 
minimum, convergence is achieved faster by using good initial guesses.  This approach was best suited to our method of spectral analysis.

\subsection{Simple Synthetic Spectrum Fitting  \label{subsec:simplefitsec}}

In order to get approximate parameters for the gas-phase lines we observed, we first fit synthetic spectra to the data using a fitting method 
commonly referred to as ``chi by eye.''  This approach allowed us to get quick initial estimates of parameters such as excitation temperature 
and column density for the various molecules that were used as inputs to a more sophisticated non-linear least squares code to obtain 
optimal values of the parameters.  The simple method also allowed us to determine the number, width and velocities of line components.

\begin{figure*}\centering
\includegraphics[angle=0,width=1.00\textwidth]{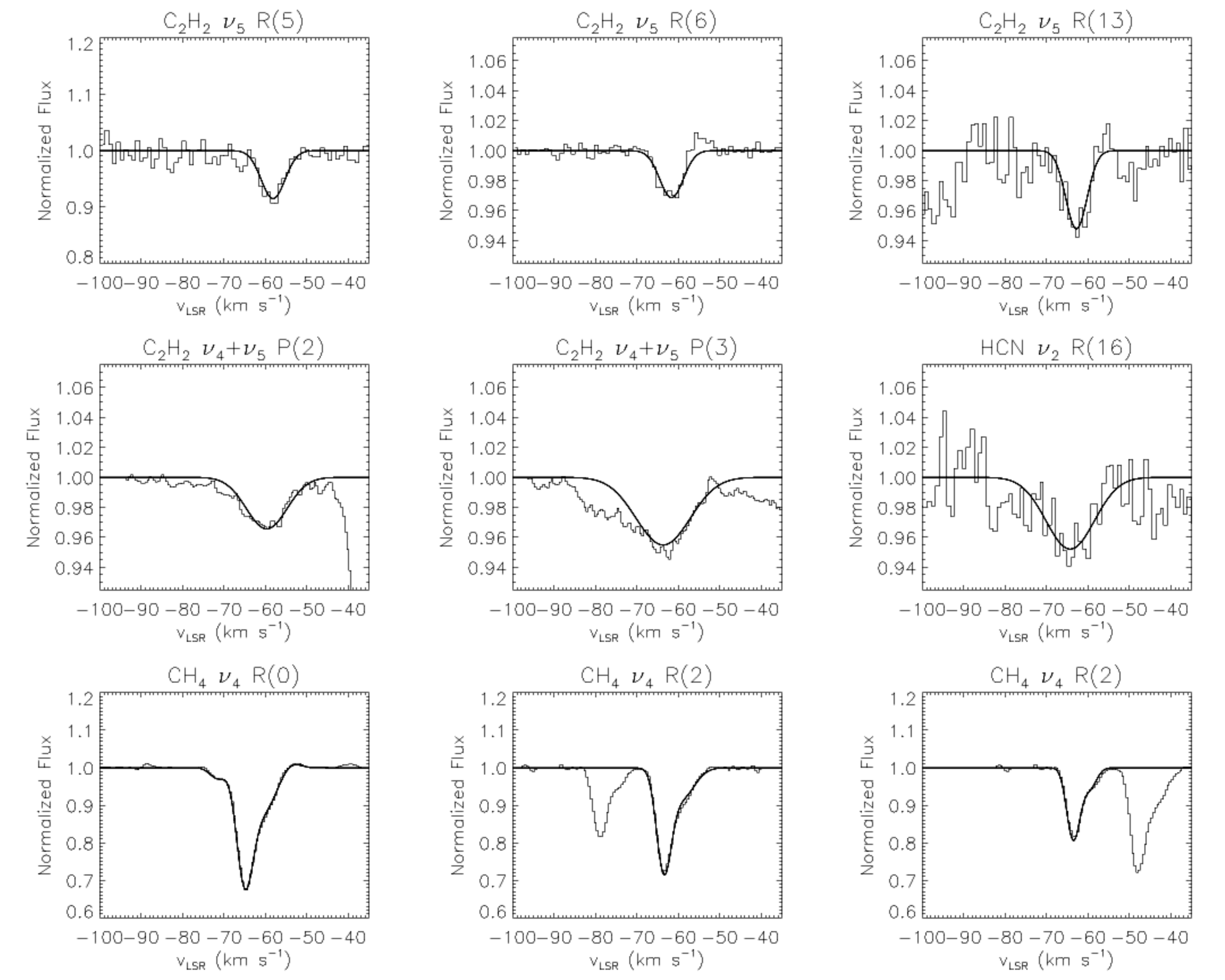}
\caption{TEXES spectra of NGC 7538 IRS 9 (thin lines) and synthetic spectra computed using the simple LTE fitting method described in 
Section~\ref{subsec:simplefitsec} (thick lines).  The continuum in each spectrum has been normalized to unity and the spectra are plotted on 
a velocity scale referred to the LSR.  Features shown are C$_2$H$_2$ ${\nu}_5$ $R$(5), $R$(6) and $R$(13);  C$_2$H$_2$ ${\nu}_{4}+{\nu}_{5}$ $P$(2) and $P$(3); HCN ${\nu}_2$ $R$(16); and CH$_4$ ${\nu}_4$ $R$(0) and the two components of $R$(2).}
\label{fits_p1}  
\end{figure*}

\begin{figure*}\centering
\includegraphics[angle=0,width=1.00\textwidth]{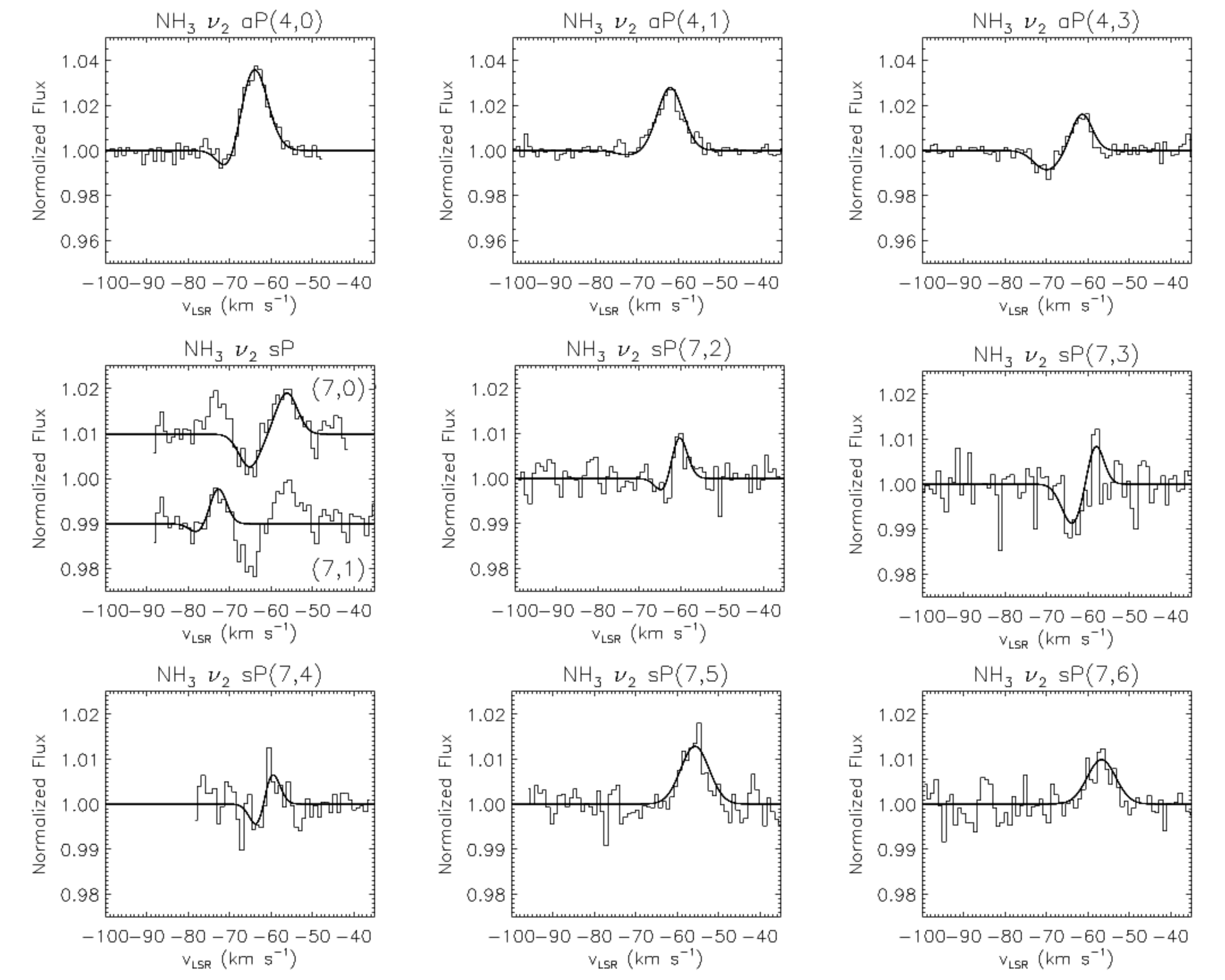}
\caption{TEXES spectra of NGC 7538 IRS 9 (continued).  Features shown are NH$_3$ ${\nu}_2$ a$P$(4,0), a$P$(4,1), a$P$(4,3), s$P$(7,0), s$P$(7,1), s$P$(7,2),  s$P$(7,3), s$P$(7,4), s$P$(7,5) and s$P $(7,6).  The s$P$(7,0) and s$P$(7,1) lines are shown together in a single panel using the velocity scale for s$P$(7,0), and are offset slightly for clarity.}
\label{fits_p2}  
\end{figure*}

\begin{figure*}\centering
\includegraphics[angle=0,width=1.00\textwidth]{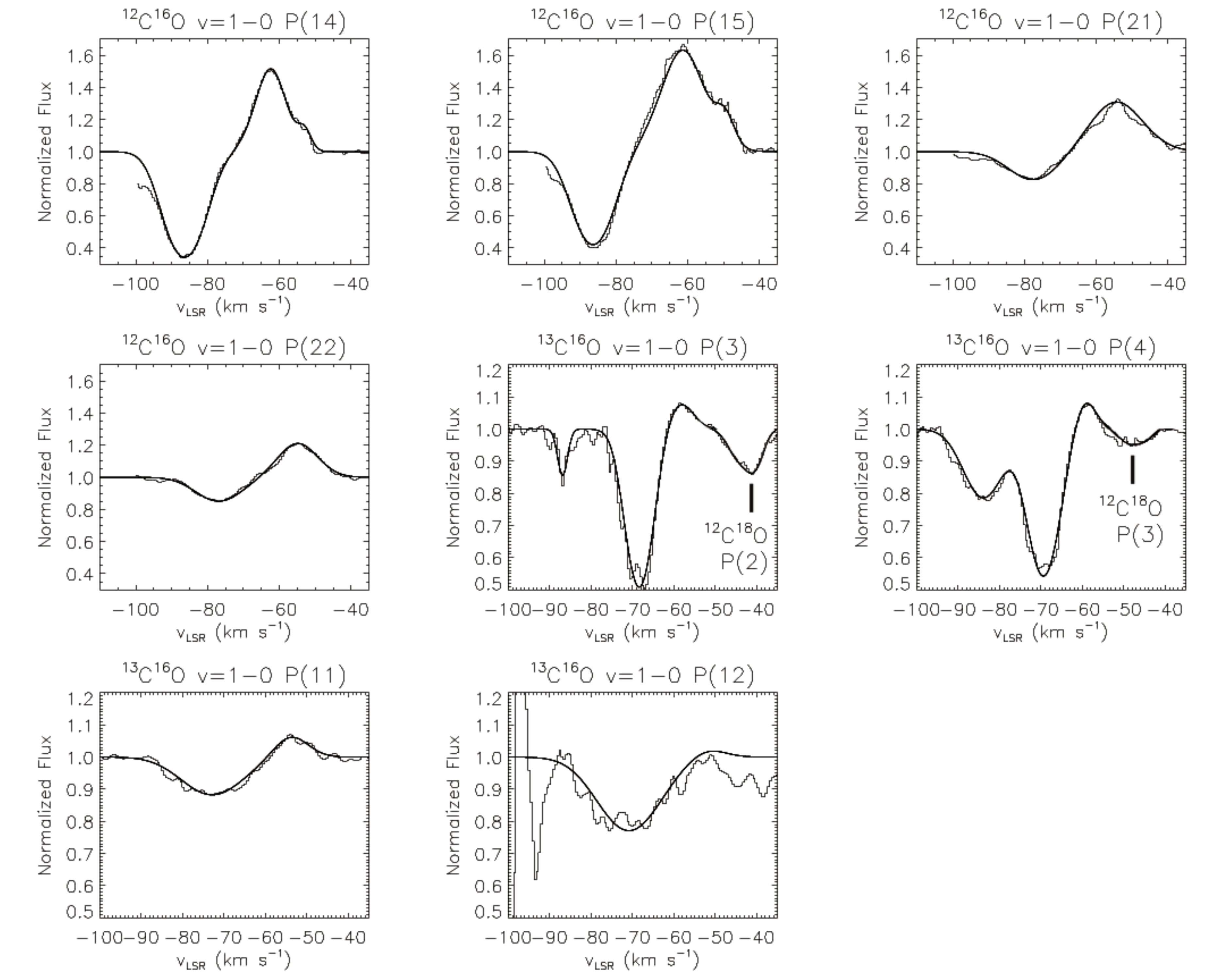}
\caption{TEXES spectra of NGC 7538 IRS 9 (continued).  Features shown are $^{12}$C$^{16}$O $v$=1-0 $P$(14), $P$(15), $P$(21) and $P$(22); $^{13}$C$^{16}$O $v$=1-0 $P$(3) / $^{12}$C$^{18}$O $v$=1-0 $P$(2); $^{13}$C$^{16}$O $v$=1-0 $P$(4) / $^{12}$C$^{18}$O $v$=1-0 $P$(3); and $^{13}$C$^{16}$O $v$=1-0 $P$(11) and $P$(12).}
\label{fits_p3}  
\end{figure*}

The equivalent width of an optically-thin line is the product of the line strength, $\alpha$, and the column density, $N$.  We obtained values of  $\alpha$ for each molecular species observed from the GEISA database\footnote{http://ether.ipsl.jussieu.fr/ether/pubipsl/geisa\_iasi\_raie\_frame\_2003\_uk.jsp} of laboratory molecular data \citep{GEISA}.  The GEISA coefficients are specified for a fiducial temperature $T_o$ = 296 K; we corrected them to an assumed temperature $T$ via the formula
\begin{equation}
{\alpha}(T) = {\alpha}(T_o)\exp\left[{{E_l \left( \frac{1}{{k}_{B}{T_o}} - \frac{1}{{k}_{B}T}\right)}}  \right] \left( \frac{T_o}{T}\right)^{\beta}
\end{equation}
where $E_l$ is the energy of the lower state of a transition, $k_B$ is Boltzmann's constant, and $\beta$ = 1 for linear molecules and $\frac{3}
{2}$ for non-linear molecules.  Here the temperature dependence of the vibrational partition function and the stimulated emission correction 
are neglected as both are of order unity.  For Gaussian lines, the line-center optical depth is calculated according to
\begin{equation}
{\tau}_l = \frac{N{\alpha}}{b{\sqrt{\pi}}}
\end{equation}
where $b$ is the Doppler $b$ parameter in wavenumber units.  The run of optical depth with frequency over the line is computed with
\begin{equation}
{\tau}(w)={\tau}_{l}\exp{\frac{-(w-{w}_{o})^{2}}{b^2}}
\end{equation}
in which $w$ is the wavenumber and $w_o$ is the line-center wavenumber.  The optical depths are converted to intensities by taking the 
exponential of ${\tau}$($w$); the continuua are normalized to unity.  We assume that the rotational states are in local thermodynamic 
equilibrium (LTE).  Our radiative transfer model also assumes pure absorption for fitting the observed absorption lines. Re-emission by the 
absorbing molecules is neglected.  Emission features are treated as separate components, which are modeled as negative absorption.  
Populations of excited vibrational states are not calculated.  Temperatures  are arbitrarily held fixed at 200 K; this value is consistent with 
temperatures around IRS 1 derived from TEXES data.  The exception was the $Q$-branch of C$_2$H$_2$ where a number of observed $J
$ states allowed the temperature to be constrained to 100 K.  Column densities and velocities derived using this method are shown in 
Table~\ref{simplefits}.  Fits to the individual lines are shown in Figures~\ref{fits_p1} through ~\ref{fits_p3}.  
\begin{deluxetable*}{ccccccccccccc}
\tabletypesize{\footnotesize}
\tablecolumns{13}
\tablewidth{0pt}
\tablecaption{Molecular absorption parameters derived from simple synthetic spectrum fitting in LTE}
\tablehead{
\colhead{Molecule/Band} & \colhead{Line} & \colhead{n\tablenotemark{a}} & \colhead{T} & \colhead{$v_1$\tablenotemark{b}} & \colhead{$v_2$} & \colhead{$v_3$} & \colhead{$b_1$} & \colhead{$b_2$} & \colhead{$b_3$} & \colhead{$N_1$\tablenotemark{c}} & \colhead{$N_2$} & \colhead{$N_3$} \\
 & & & \colhead{(K)} & \colhead{(km s$^{-1}$)}  & & & \colhead{(km s$^{-1}$)}  & & & \colhead{(cm$^{-2}$)}  & \\
}
\startdata
C$_2$H$_2$ ${\nu}_5$& $R$(1) & 1 & 100 & -62.0       &      -- & -- & 3.3 &   -- & --   & 1.0(15) & -- & --  \\	
          & $R$(5) & 1 & 100 & -62.0       &      -- & -- &  2.4 &   --   & -- & 4.5(14) & -- & -- \\
 	& $R$(6) 	& 1 & 100 & -65.5 & -- 	& --   &      2.4 &  --  & -- &      4.5(14) & -- & -- \\
 	& $Q$ branch & 1 & 100     &     --  & --  & -- & 1.0 & -- & -- &  2.0(15) & --  & -- \\
C$_2$H$_2$ ${\nu}_{4}+{\nu}_{5}$ & P(2) & 1 & 100 & -61.0 & -- & -- & 6.8   & -- & -- & 2.5(16)   & -- &      --    \\
         & P(3) & 1 & 100 & -62.0 & --   & -- & 5.7 & --    & -- &   1.0(16) & -- & -- \\
HCN ${\nu}_2$ & $R$(16) & 1 & 200 & -64.5 & -- & -- & 7.9 & -- & -- & 1.2(16) & -- & -- \\
$^{12}$CH$_4$ ${\nu}_4$ & $R$(0)\tablenotemark{d} & 4 & 200 & -65.0 & -54.0  & -61.0 &  2.3 & 2.7 & 4.6 & 7.5(16) &  -4.0(15) & 6.0(16) \\
 	& $R$(2) 	& 2 & 200 & -65.5 & -61.5   & -- & 2.3 & 2.5   & --   &      6.0(16) & 3.3(16) & -- \\
	 & $R$(2) 	& --  & -- &  -- & -- & -- & 2.5 & 2.3 & -- & 6.3(16) & 1.5(16) & -- \\
$^{14}$NH$_3$ ${\nu}_2$ & a$P$(4,0) 	& 2 & 200 & -64.0 & -70.0   &      -- & 4.6 & 3.9     & -- &      -2.5(15) & 5.0(14) & --\\
         & a$P$(4,1) 	& 2 & 200 & -62.0 & -72.0   & -- &      5.3 & 3.9     & -- &      -3.8(15) & 2.5(14)  & -- \\
 	& a$P$(4,3) 	& 2 & 200 & -62.0 & -70.0   & -- &      2.8 & 3.9     & -- &      -1.0(15) & 1.0(15) & -- \\
 	& s$P$(7,0) 	& 2 & 200 & -60.0 & -69.0   & -- &      2.9 & 2.9     & -- &      -1.5(15) & 1.2(15) & -- \\
 	& s$P$(7,1) 	& 2 & 200 & -62.0 & -70.0   & -- &      1.8 & 4.0     & -- &      -2.0(15) & 7.5(14)   & -- \\
 	& s$P$(7,2) 	& 2 & 200 & -61.0 & -62.5   & -- &      3.3 & 2.9     & -- &      -6.0(15) & 4.5(15) & -- \\ 
 	& s$P$(7,3) 	& 2 & 200 & -58.5 & -64.0   & -- &      1.8 & 2.9     & -- &      -1.2(15) & 1.5(15) & -- \\
 	& s$P$(7,4) 	& 2 & 200 & -60.5 & -63.0   & -- &      1.8 & 1.8     & -- &      -2.5(15) & 2.0(15) & -- \\ 
 	& s$P$(7,5) 	& 1 & 200 & -56.0 & -- 	  & -- & 4.4 & --        & -- &      -7.0(15) & -- & -- \\ 
 	& s$P$(7,6) 	& 1 & 200 & -57.0 & -- 	  & -- & 4.4 & --        & -- &      -4.0(15) &--  & -- \\ 
$^{12}$C$^{16}$O v=1-0 & $P$(14) & 3 & 200 & -55.0 & -64.5 & -88.5 & 2.4 & 6.5 & 7.2 & -1.0(16) & -1.1(17) & 3.0(17) \\
         & $P$(15) 	& 3 & 200 & -51.5 & -63.5 & -88.5 & 3.6 & 9.4 & 7.2 & -2.5(16) & -2.2(17) & 3.0(17) \\
         & $P$(21) 	& 2 & 200 & -54.5 & -77.5   & -- &      10.9 & 10.9    & -- &      -8.0(17) & 5.5(17) & -- \\
 	& $P$(22) 	& 2 & 200 & -55.0 & -77.0   & -- &      8.0 & 10.2    & -- &      -6.0(17) & 6.0(17) & -- \\
$^{13}$C$^{16}$O v=1-0& $P$(3) & 3 & 200 & -60.5 & -70.5 & -89.0 & 4.3 & 4.3 & 1.4 & -9.0(17) & 8.0(18) & 7.0(17) \\
	& $P$(4)		& 3 & 200 & -61.5 & -71.5 & -86.0 & 3.6 & 5.0 & 6.5 & -7.0(17) & 6.8(18) & 3.4(18) \\
	& $P$(11) 	& 2 & 200 & -54.0 & -75.0  & -- & 5.8 & 10.2  & -- &  -8.0(17) & 2.7(18) & -- \\
	& $P$(12) 	& 2 & 200 & -54.0 & -73.0 & -- & 5.9 & 10.2 & -- & -3.5(17) & 6.5(18) & -- \\
$^{12}$C$^{18}$O v=1-0& $P$(2) & 2 & 200 & -68.0 & -72.0 & -- & 2.9 & 3.6 & -- & 7.5(18) & 7.5(18) & -- \\ 
	& $P$(3)		& 2 & 200 & -66.0 & -71.0 & -- & 4.3 & 4.3 & -- & 3.0(18) & 3.0(18) & -- \
\enddata
\tablenotetext{a}{Number of velocity components fit.}
\tablenotetext{b}{All velocities are referred to the LSR.}
\tablenotetext{c}{Column densities are given in the form $A$($B$) = $A$$\times$10$^{B}$.  Negative values of $A$ indicate components in emission.  ``Column density'' includes all rotational states, assuming $T$ = 200 K.}
\tablenotetext{d}{The best fit to this line required four components.  The remaining component not in the table is $v_4$ = -71.5 km s$^{-1}$, $b_4$ = 2.3 km s$^{-1}$, $N_4$ = 5.0(15)cm$^{-2}$}.
\label{simplefits}
\end{deluxetable*}

\subsection{${\chi}^2$ Minimization Spectral Fitting\label{ss:FITSPEC}}

Our spectral fitting code, \tt FITSPEC\rm,  uses the Levenberg-Marquardt  method of inverting the second derivative matrix to minimize ${\chi}
^2$ as in \citet{Bevington} for up to four unique absorbing or emitting units, or ``clouds,'' along the line of sight.  The 1-D radiative transfer 
model assumes rotational LTE of all molecules with a single excitation temperature.  As in our ``by eye'' fitting, re-emission by absorbing 
molecules is neglected and emission features are modeled as negative absorption.  The total number of free fitting parameters depends on 
the number of molecules in the fit and the number of echelon orders in a particular spectral setting.  Although there is a large number of free 
parameters in the fit, most of them are ``uninteresting,'' such as baseline coefficients.  The ``interesting'' parameters in the fit are Doppler 
shift, Doppler $b$ parameter, column density, temperature, and covering factor for each cloud and for each molecule.  The details of the 
program are discussed in \citet{Claudia}.  For our observed molecules, we chose a number of absorbing clouds consistent with the number 
of velocity components seen in the TEXES spectra, allowing \tt FITSPEC \rm to independently determine their parameters.  When available, 
we included data indicating the non-detection of certain lines, such as high-$J$ C$_2$H$_2$ near 13 ${\mu}$m, to better constrain the fit 
parameters.

The success of \tt FITSPEC \rm requires accurate molecular transition data to compute synthetic spectra against which to fit the data.  We 
relied on the calculations in \citet{Evans91}, giving information including the line center, the energy of the transition either in units of $J$($J$
+1) (which is then multiplied by the molecule's rotational constant $B$) or $E$ in units of cm$^{-1}$, and the rotational line strengths (H\"{o}nl-London factors) in units of cm$^{-1}$ (10$^{16}$ cm$^{-2}$)$^{-1}$.  The program uses guesses of the starting parameter values to 
compute the initial model spectrum, and then iterates the fit by varying those values until a convergence criterion is satisfied.  The 
requirement is
\begin{equation}
\sum_i \left (\frac{{\Delta}p_i}{{\sigma}_{p_i}}\right )^{2} \leq 0.01
\end{equation}
where ${\Delta}p_i$ is the change in the value of parameter $i$ and ${\sigma}_{p_i}$ is the uncertainty in the parameter $i$, allowing all 
other parameters to vary.

The output of the program consists of the best-fit parameter values, uncertainty estimates and a synthetic spectrum having the properties of 
the best fit.  The intensity of the synthetic spectrum is calculated by taking $I$ = $I_o$$e^{-{\tau}}$ of each absorbing unit along the line of 
sight and adding any optically thin emission as
\begin{equation}
I_{obs} = {I_0}\prod_{n=1}^N [1-{C_n}(1-e^{-{{\tau}_n}(\nu)})] +S_{em} C_{em} \tau_{em}(\nu)
\label{iprod}
\end{equation}
where $N$ is the number of absorbing clouds in the fit, $I_0$ is the intensity of radiation from the background continuum source (set to unity 
since the spectra are normalized), $C_n = \Omega_n / \Omega_0$ and $\tau_n$ are the covering factors and optical depths of the 
absorbing clouds, $S_{em}$ and $C_{em} = \Omega_{em} / \Omega_{0}$ are the source function and covering factor of the emitting 
material, and $\Omega_0$ is the solid angle of the continuum source.  The right-hand term represents an optically-thin emitting cloud; since 
it only depends on the product of $S_{em}$ and $C_{em}$, we scale $C_{em}$ by $S_{em} / I_0$ and drop $S_{em}$ from the formula.  The 
product term in Equation~\ref{iprod} assumes that the absorbing clouds overlap randomly.  That is, if two clouds each have covering factors 
of 0.1, 1/10 of each lies along the line of sight to the other.

\citet{Claudia} used the same approach and note this results in greater absorption in saturated lines than would result if the optical depths 
were added first and then the transmission spectrum were calculated from the optical depth spectrum.  The resulting fits to the data are 
better than those obtained by summing the optical depths first and then taking the exponential.  This fitting approach, however, is not without 
drawbacks.  Emission lines may not be thermally excited, but rather could be due to resonantly scattered light; we have no way of 
discriminating between these processes.  We assume that upward transitions in the molecules we observe are not followed by re-emission 
of that radiation; in reality, most absorptions should be followed by re-emission, although in a direction uncorrelated with the incoming 
radiation, so the extent to which re-emission fills in the absorption depends on the distribution of the re-emitting gas around the continuum 
source.

The results of our \tt FITSPEC \rm runs on the IRS 9 data are summarized in Table~\ref{FITSPEC_results} where they are compared with 
results from fitting TEXES data of IRS 1 with the same procedure in \citet{Claudia}.  
\setlength{\tabcolsep}{0.02in} 
\begin{deluxetable}{cccccc}
\tabletypesize{\footnotesize}
\tablecolumns{6}
\tablewidth{0pt}
\tablecaption{Molecular absorption parameters determined from ${\chi}^{2}$ minimization spectral fitting}
\tablehead{
\colhead{Species} & \colhead{$v_{LSR}$} & \colhead{T$_{ex}$} & \colhead{$b$} & \colhead{$N(X)$\tablenotemark{a}} & \colhead{C} \\
& \colhead{(km s$^{-1}$)} & \colhead{(K)} & \colhead{(km s$^{-1}$)} & \colhead{(10$^{16}$ cm$^{-2}$)} & \colhead{\%}}
\startdata
\cutinhead{IRS 9 (this work)}
C$_2$H$_2$ 	& -58.9(1.4) & 195(57) & 4.0(1.3) & -0.2(0.1) & 52(189)\tablenotemark{b} \\
 	& -63.9(1.0) & 205(49) & 4.0(1.1) & 2.3(0.7) & 16(9)	\\	
HCN & -58.7(1.7) & 200(68) & 4.0(1.3) & 0.5(0.7) & 52(193)\tablenotemark{b} \\
	& -63.8(2.9) & 204(76) & 4.0(1.4) & 3.9(4.3) & 11(6) \\
NH$_{3}$\tablenotemark{a} & -57.3(0.5) & 365(47) & 5.1(0.7) & -1.1(0.1) &	 52(195)\tablenotemark{b} \\
	& -61.6(0.6)	& 201(43)	& 5.9(1.1) & 4.8(0.8) & 8(8) \\
HNCO\tablenotemark{c} & -58.9(0.0) & 195(0) & 4.0(0.1) & 0.1(-0.5) & 52(0)	\\ 	
	& -63.9(10.0) & 205(50) & 4.0(1.0) & 0.9(-1.6) & 16(10) \\
\cutinhead{IRS 1 \citep{Claudia}} 
C$_2$H$_2$ 	& -55.7(0.3) & 225(20) & 0.6(0.1) & 3.0(0.6) & 6(0.6) \\
			& -59.4(0.3) & 191(10) & 0.6(0.1) & 2.8(0.2) & 24(1) \\
HCN			& -56.3(0.3) & 256(30) & 0.8(0.1) & 5.6(2.8) & 8(1) \\
			& -60.0(0.3) & 456(127) & 1.0(0.4) & 1.3(1.2) & 68(50) \\
NH$_{3}$  	& -57.3(0.3) & 278(31) & 0.9(1.0) & 5.2(1.5) & 9(1) \\
			& -60.1(0.3) & 248(32) & 0.4(0.1) & 2.8(0.8) & 16(4) \\
HNCO 		& -57.2(0.3) & 319(27) & 1.8(0.2) & 0.4(0.3) & 4(0.3) \\			
                            & -60.2(0.2) & 171(79) & 0.7(0.4) & 0.1(0.3) & 97(11) \\
\enddata
\tablecomments{Quantities in parentheses are the 1$\sigma$ uncertainties on the adjacent figures in the same units unless otherwise indicated.}
\tablenotetext{a}{Negative column densities indicate components in emission.  Column densities for the emission components are $C_{em}\tau_{em}$($\nu$) converted to $N$ and do not represent a real column densities.}
\tablenotetext{b}{Unable to be constrained in the fit, assuming lines are optically thin.}
\tablenotetext{c}{All HNCO parameters other than the column density were fixed equal to those of C$_2$H$_2$ because an insufficient number of HNCO lines were observed to constrain them.}
\label{FITSPEC_results}
\end{deluxetable}
\setlength{\tabcolsep}{6pt} 

\subsection{Column Density Upper Limits For Non-Detections\label{ss:upper}}

We calculate column density upper limits from the lines that were not detected in the spectrum of IRS 9 by convolving the data with a 
Gaussian line shape function whose properties are consistent with those of detected lines.  We begin by  assuming a normalized spectrum 
containing a single absorption line has the functional form 
\begin{equation}
F_\nu = 1 - ae^{-x^{2}/b^{2}}
\end{equation}
where $a >$ 0 is the depth of the line and $b$ is the Doppler $b$ parameter.  The observed spectrum is convolved with a Gaussian of the 
form $ne^{-x^{2}/b^{2}}$ where $n$ is a normalization factor here chosen to be 1/${\sqrt{\pi}}b$ such that the integral of the Gaussian is 
unity.  The convolution is given by
\begin{equation}
F_{\nu} \otimes ne^{-x^{2}/b^{2}} = \int F_{\nu} ne^{-x^{2}/b^{2}} d\nu,
\end{equation}
where $x={\nu}-{\nu}^{\prime}$.  The upper limit of the equivalent width, $W_{max}$, of a line not detected is proportional to the peak value 
of the convolved spectrum according to
\begin{equation}
W_{max} = {\sqrt{\pi}}Pb
\end{equation}
in which $P$ is the peak.  We used $b$ = 3.5 km s$^{-1}$ in making the calculations, a value consistent with those derived by spectral fitting 
methods described previously.  The column density limit for species $X$ was then computed from
\begin{equation}
N(X) = \frac{W_{max}}{\alpha}
\end{equation}
where $\alpha$ is the line strength.  We used the GEISA line strengths where available and corrected them to a uniform temperature of 200 
K as described previously.  Results for our non-detections are given in Table~\ref{nondetects}.
\begin{deluxetable}{cccc}
\tabletypesize{\footnotesize}
\tablecolumns{5}
\tablewidth{0pt}
\tablecaption{3${\sigma}$ upper limits to column density of molecules in the upper state for selected non-detections}
\tablehead{
\colhead{Species/Mode} & \colhead{Line} & \colhead{$n$\tablenotemark{a}} & \colhead{$N_{limit}$(X) (cm$^{-2}$)}}
\startdata
$^{13}$CH$_4$ ${\nu}_4$\tablenotemark{b}  		& $R$(1) & 1 & 4.1${\times}$10$^{16}$ \\
  							& $R$(4) & 4 & 3.1${\times}$10$^{16}$ \\
OCS ${\nu}_2$ 				& $P$ branch & 34 & 2.9${\times}$10$^{15}$ \\
HCN ${\nu}_2$ 				& $R$(15) & 1 & 1.8${\times}$10$^{16}$ \\
							& $R$(22)	& 1 & 8.5${\times}$10$^{16}$ \\
 							& $R$(23)	& 1 & 1.1${\times}$10$^{17}$ \\
C$_2$H$_2$ ${\nu}_5$ 			& $R$(12)	& 1 & 8.5${\times}$10$^{15}$ \\
							& $R$(21)	& 1 & 9.4${\times}$10$^{15}$ \\
 							& $R$(22)	& 1 & 3.5${\times}$10$^{16}$ \\
C$_2$H$_2$ $2{\nu}_{5}-{\nu}_{5}$ & $Q$ branch & 15 & 2.1${\times}$10$^{17}$ \\
\enddata
\tablecomments{A uniform temperature of 200 K was assumed for all species.}
\tablenotetext{a}{Number of lines/components in the attempted fit.}
\tablenotetext{b}{The given values of $N$($^{13}$CH$_4$) are for the isotopologue only rather than the total CH$_4$ column density.}
\label{nondetects}
\end{deluxetable}

We derived upper limits for all lines of a particular molecular species available in a given spectral setting, as indicated in the table by the 
number of lines parameter $n$.  In cases where $n$ $>$ 1, the equivalent width and column density upper limits given are simple averages.  
All limits are given with respect to the total number of molecules along the column, irrespective of isotopic composition except for $^{13}$CH$_4$; here we have divided the derived column density by the terrestrial $^{12}$C/$^{13}$C = 89 ratio built into GEISA to recover the 
column density specifically for $^{13}$CH$_4$.

\subsection{An Upper Limit to the Ionizing Flux in IRS 9\label{subsec:neflux}}

The $^{2}P_{1/2}$ $\to$ $^{2}P_{3/2}$ fine structure line of [\ion{Ne}{2}] at 780.42 cm$^{-1}$ is often used as a diagnostic of ionized gas at 
our wavelengths. The non-detection of this line in the spectrum of IRS 9 places a constraint on the amount of ionizing flux being emitted by 
the embedded central source.

The Lyman continuum luminosity ($N_{Lyc}$) can be calculated from the [\ion{Ne}{2}] luminosity since both are proportional to the square of the density integrated over volume.  The [Ne II] volume emissivity is given by 
\begin{equation}
j_{[NeII]} = n_e n_{Ne^+} q_{lu} h \nu {n_c \over {n_e + n_c}}
\end{equation}
where $n_e$ is the electron density, $n_{Ne^{+}}$ is the density of ionized neon,  $q_{lu}$ is the rate coefficient for collisional excitation 
($q_{lu} h \nu = 8.4 \times 10^{-22}$ erg s$^{-1}$ cm$^{3}$), and $n_c$ = $A_{ul}$/$q_{ul}$ (= 4.9 $\times$ 10$^5$ cm$^{-3}$ at an electron 
temperature  $T_e = 10^4$ K) is the critical density in the two-level approximation, using the values in \citet{osterbrock}.  The hydrogen 
recombination rate per unit volume is given by $n_{recomb} = \alpha_B n_e n_p$ where $\alpha_B$ (= 2.6 $\times 10^{-13}$ cm$^3$ s$^
{-1}$) is the recombination rate for ``case B'' recombination, and $n_p$ is the proton density.  The ratio of $N_{Lyc}$ to the [\ion{Ne}{2}] luminosity is then given by 
\begin{equation}
\frac{N_{Lyc}}{L_{[NeII]}} = \frac{n_{recomb}}{j_{[NeII]}} = {n_{H^+} \over n_{Ne^+}} {\alpha_B \over {q_{lu} h \nu}}
\end{equation} 
assuming $n_e \ll n_c$.  The Ne$^{+}$/H$^{+}$ ratio is given by the product of the Ne abundance and the singly-ionized fraction for Ne, $f_
{+}$.   We take the value of $f_{+}$ = 0.3 from Figure 3 in \citet{Ho07} using for the radiation temperature that of a ZAMS star with IRS 9's 
luminosity ($T_r$ = 26200 K) and a ratio of the dilution factor for blackbody radiation, $\Gamma$, to the electron density of $\Gamma$/$n_e
$ = 10$^{-19}$ cm$^{3}$.  From this we adopt a Ne$^{+}$ / H$^{+}$ ratio to be $3 \times 10^{-5}$.

Without knowing the width of the [Ne II] line, it is difficult to place a strong upper limit on its flux, but an emission equivalent width of $W_{\nu} 
= 0.01$ cm$^{-1}$ can be ruled out unless the line is wider than 0.2 cm$^{-1}$, or 77 km s$^{-1}$.  This corresponds to a line flux limit of $F_
{[NeII]} < 2.0 \times 10^{-13}$ erg s$^{-1}$ cm$^{-2}$ and a luminosity limit of $L_{[NeII]} = 4 \pi d^2 F_{[NeII]} < 1.7 \times 10^{32}$ erg s
$^{-1}$.  We then calculate $N_{Lyc} < [3.3 \times 10^4 \times 2.6 \times 10^{-13} / 8.4 \times 10^{-22}] \times 1.7 \times 10^{32} = 1.7 \times 
10^{45}$ photons s$^{-1}$.  For reference, \citet{Tielens} gives the Lyman continuum luminosity of a main sequence B0 star as $N_{Lyc} = 
1.4 \times 10^{48}$ photons s$^{-1}$.  Apparently IRS 9 emits a small fraction of the Lyman continuum luminosity of a main-sequence star of 
its bolometric luminosity.  The radiation environment of IRS 1 is considerably more intense, by comparison.  From infrared and millimeter 
data, the exciting source is inferred to be of spectral type earlier then O7.5, total luminosity $L$ $>$ 8 $\times$ 10$^4$ $L_{\odot}$, and 
Lyman continuum luminosity $N_{Lyc} >$ 10$^{48}$ s$^{-1}$ \citep{Hackwell82,Akabane01,Lugo04}.  The inferred radiation field of IRS 1 is more intense than that of IRS 9 and results in a different set of conditions under which chemistry proceeds.

\section{Interpretation of the Spectroscopic Results\label{sec:interp}}

\subsection{Assumptions and Caveats\label{subsec:caveats}}

Interpretation of our results requires caution, given the underlying assumptions of the radiative transfer model in \tt FITSPEC\rm.  The 
program assumes a single continuum source behind all absorbing material and a constant temperature characterizing the material in each 
of up to four absorbing ``clouds'' along the line of sight.  We also add an optically-thin emitting cloud, but neglect re-emission by the 
absorbing gas and absorption by the emitting gas.  The column densities we report do not represent the entire column down to the 
embedded luminosity source, but rather the column only to an effective ``dust photosphere'' interior to which the continuum optical depth is 
large at a given wavelength.  Consequently, our observations at 5 $\mu$m and 8 $\mu$m likely probe to greater depth than those at 13 $
{\mu}$m.  The optical depth may be higher at longer wavelengths due to silicate opacity, and the colder, outer layers of the dusty region are 
likely too cool to emit significantly at 5 $\mu$m.  Comparison of results at different wavelengths is therefore difficult.  We also did not take into 
account emission by molecules following absorption of photons, except as approximated by our ``negative absorption'' method of treating 
emission lines.  Furthermore, we ignored emission following radiative excitation.  We assumed either pure absorption or optically-thin 
emission.  However, the neglect of emission from a radiatively excited molecule should not be valid; at any plausible gas density the 
collisional de-excitation rate should be far below the radiative rate.  Critical densities for vibrational transitions are typically greater than 
$10^{12}$ cm$^{-3}$, and almost all absorption events should be followed by emission, not collisional de-excitation.  For two of our 
molecules, NH$_3$ and C$_2$H$_2$, we were able to observe lines of more than one vibrational band or rotational branch, and the 
comparison of the observed line depths provides evidence that re-emission following absorption is an important effect. 

For NH$_3$, we observed weak emission in $P$-branch lines at wavelengths near 12 $\mu$m and possibly weak absorption in $Q$-
branch lines near 10.5 $\mu$m.  $R$-branch lines were not observed due to the silicate and telluric O$_3$ absorption at their wavelengths.  
We explain the differences between the $P$- and $Q$-branch lines based on the differing continuum radiation fields due to their differing 
proximity to the silicate dust feature.  The NH$_3$ $R$ branch is centered on the 9.7 $\mu$m silicate feature, the $Q$ branch falls on its 
long wavelength shoulder, and the $P$ branch is mostly between the 9.7 and 18 $\mu$m features.  If the radiation field seen by a molecule 
is due to silicate emission either because the molecule is exposed to radiation from an optically thin external dust cloud or because it is 
immersed in a cloud of warm silicate dust, there will be more upward radiative transitions (i.e. more absorption) in the $Q$ and $R$ 
branches than in the $P$ branch.  The branching ratio between $P$-, $Q$-, and $R$-branch transitions from a given vibrationally-excited 
state depends on the $K$ rotational quantum number, but is similar in the three bands.  Consequently, if all absorption events are followed 
by an emission, absorptions will outnumber emissions in the $R$ branch and emissions will outnumbers absorptions in the $P$ branch, 
resulting in net emission in the $P$-branch lines, as we observe.  In Section~\ref{subsec:nh3scheme} we describe a simple model involving 
purely radiative transitions in NH$_3$ incorporating upward and downward line strengths and assuming that the incident radiation field is 
proportional to the strength of the silicate feature at the wavelengths of the various lines.  The model is time-independent and assumes the 
NH$_3$ gas is optically thin.  More detailed and quantitative models will follow in a future paper.

\subsection{C$_2$H$_2$, HCN, HNCO, OCS, and CO}

We find that excitation temperatures for various molecules are basically consistent toward IRS 9 and IRS 1 with the exception of the hotter of 
the two NH$_3$ components toward IRS 1.  That value of $T_{ex}$ has a rather large uncertainty so it may in fact lie closer to $\sim$250 K.  
The column densities are also similar in the two sources, though again with particular exception in the case of HCN.  \citet{Claudia} had the 
benefit of more observed HCN lines, but we find a lower column toward IRS 9.  Most striking is the disagreement between the two sources in 
terms of the intrinsic line widths; although the thermal broadening should be similar,  turbulence or systematic velocity variations along the 
line of sight to IRS 9 are greater than toward IRS 1.   At the least, we can say that the spectral lines observed in the direction of IRS 9 are 
generally less saturated than those seen toward IRS 1, and this has important consequences for the accuracy of the IRS 9 column densities 
we obtain.

An example in which saturation of lines becomes important is CO.  \citet{Claudia} derived fractional abundances relative to CO for a variety 
of molecules seen in absorption toward IRS 1, attempting to compensate for badly saturated $^{12}$CO lines by using the value of $N$($^
{13}$CO) in \citet{Mitchell90} and adopting $^{12}$C/$^{13}$C = 45 on the basis of their observations of C$_2$H$_2$.  The value of $N$($^
{13}$CO) for IRS 9 reported by Mitchell et al. was based on their measurement of $^{12}$CO lines and assumed the terrestrial ratio $^
{12}$C/$^{13}$C =  89.  \citet{BoogertBlake04} confirm the saturation of the $^{12}$CO lines observed toward IRS 9.  From observations of 
lines of $^{13}$CO and $^{12}$C$^{18}$O in upper $J$ levels of 7-15 they derive an optical depth ${\tau}_{12}$ = 13 $\pm$ 4 assuming $^
{12}$C/$^{13}$C =  80 from \citet{Boogert02}.  However, they argue that reliable values for $N$($^{12}$CO) can be obtained, since the $^
{13}$CO lines remain optically thin; they quote a value $N$($^{12}$CO) =  (3.2$\pm$1.0)$\times$10$^{18}$ cm$^{-2}$.  While we may 
argue with their choice of $^{12}$C/$^{13}$C and with their conclusion that the $^{13}$CO lines are optically thin, their value of the CO 
column density is probably more realistic for IRS 9 than that obtained from the lower-resolution observations of \citet{Mitchell90}.  
Abundance ratios for some of the molecules we observed with TEXES, relative to this column of CO, are shown in Table~\ref{abundances}.
\begin{deluxetable*}{cccccc}
\tabletypesize{\footnotesize}
\tablecolumns{5}
\tablewidth{0pt}
\tablecaption{Abundances in IRS 9 and IRS 1 with respect to CO and H$_2$}
\tablehead{
& \multicolumn{2}{c}{IRS 9 (this work)} & & \multicolumn{2}{c}{IRS 1 \citep{Claudia}} \\
\cline{2-3} \cline{5-6} \\
 \colhead{Molecule} & $N$(X)/$N$(CO) & $N$(X)/$N$(H$_2$) & & $N$(X)/$N$(CO) & $N$(X)/$N$(H$_2$)
 }
\startdata
C$_2$H$_2$ 	&	7.7(-3)	&	3.3(-7)	&	&	5.7(-3) & 7.7(-7) \\
HCN 	&	1.4(-2)	&	5.8(-7)	&	&	6.9(-3) & 9.2(-7) \\
NH$_{3}$	&	1.9(-2)	&	7.9(-7)	&	&	3.6(-2) & 4.8(-6) \\
CH$_4$ & 9.9(-2) & 4.2(-6) &  & 3.6(-2) & 4.8(-6) \\
HNCO	&	$<$3.0(-3)	&$<$1.3(-7)	&	&	5.0(-4) & 6.0(-6) \\
\enddata
\tablecomments{All values are given in the notation $A$($B$) = $A$ $\times$ 10$^{B}$.  The column densities toward IRS 1 adopted for CO and H$_2$ were $N$($^{12}$CO) = 1$\times$10$^{19}$ cm$^{-2}$ \citep{Claudia} and $N$(H$_2$) = 7.5$\times$10$^{22}$ cm$^{-2}$ \citep{Willner82}.  For IRS 9 the $^{12}$CO column density used was $N$($^{12}$CO) = 3.2$\times$10$^{18}$ cm$^{-2}$ \citep{Boogert02} and the H$_2$ column density was the same as that used by \citet{Claudia}.}
\label{abundances}
\end{deluxetable*}

Similarly, an estimate of the abundance of various molecules with respect to H$_2$ can be made given a measurement of the quantity of 
dust along the line of sight and a value for the gas-to-dust ratio in a typical protostellar envelope.  Knez et al. adopted an H$_2$ column 
density toward IRS 1 of 7.5$\times$10$^{22}$ cm$^{-2}$ based on the 9.7 $\mu$m optical depth measured by \citet{Willner82} to calculate 
fractional abundances of other molecules.  Using the same method, we derive $N$(H$_2$) = 5.2 $\times$ 10$^{22}$ cm$^{-2}$ toward IRS 
9 and present the resulting abundances in Table~\ref{abundances}.  The pattern of relative abundances differs between the objects 
according to both the molecular species and whether the abundance of a given species is referred to CO or to H$_2$.  The latter 
observation indicates that either the abundance ratio of CO to H$_2$ or the gas-to-dust ratio varies between objects.  

The assumptions made in calculating the values in Table~\ref{abundances} -- that the CO column density is comparable along both lines of 
sight and that the dust distribution and gas-to-dust ratio are similar in both cases -- may not be unreasonable given that both objects are 
along nearby sightlines toward the same molecular cloud.  Given this caveat, it is worth noting that the fractional abundances of all 
molecules relative to both CO and H$_2$ are broadly consistent with each other, except for HNCO with the values given for IRS 9 properly 
unconstrained by \tt FITSPEC \rm because no HNCO lines were detected.  C$_2$H$_2$, HCN, and NH$_3$ are generally less abundant in 
IRS 9 than IRS 1, although this is only a suggestion in light of the assumptions. 

Some kinematic information about IRS 9 is available in the line shapes.  In combination with the varying effective depths to which sightlines 
probe at different wavelengths, line shapes give a sense of the distribution of various molecules along the line of sight.  The P Cygni line 
profiles of our CO observations, for example, clearly indicate an outflow known previously.  CH$_4$ absorption toward IRS 9 was known 
previously as well, but the TEXES data reveal the structure of the lines in more detail than the Keck+NIRSPEC spectra of \citet
{BoogertBlake04} and the IRTF+Irshell spectra of \citet{Lacy91}.

In contrast to the circumstellar disk model of IRS 1 presented in \citet{Claudia} based on TEXES observations, we cannot conclusively 
establish the presence of a disk in IRS 9.  However, our fitting program preferred relatively small covering factors, similar to those in the best-fit synthetic spectra for the lower velocity component of IRS 1.  Knez et al. interpret this as the signature of a near edge-on disk, acting as a 
source of continuum radiation that is only partially covered by the absorbing material.  Given the high-velocity outflow in IRS 9 reported by 
\citet{Mitchell91}, we might rather be observing an accreting system nearly face-on in which the outflow both sweeps out a region near the 
continuum source and entrains material into the flow that only partially covers the source.  This picture would account for the methane 
observations  of \citet{BoogertBlake04} in which cold, solid-phase methane is seen in the outer envelope, becomes depleted in the outflow, 
and is seen in the gas phase in the inner envelope, all along a single sightline.  We develop this model further in Section~\ref
{subsec:toymodel}.

Non-detections of molecules also furnish some interpretable information.  Carbonyl sulfide (OCS) is a molecule that commonly occurs in 
ices along sightlines through molecular clouds and indicates the presence of sulfur-bearing precursor species such as H$_2$S and SO$_2$ \citep{OCS}.   \citet{Gibb04} detected solid-phase OCS toward IRS 9 with \it ISO\rm, finding a column density of (5.5$\pm$4.4) $\times$ 
$10^{15}$ cm$^{-2}$, and \citet{Claudia} speculate that gas-phase OCS should be observable toward IRS 9 as indicated in some chemical 
models.  We find an upper limit of $N$(OCS) $\leq$ 2.9 $\times$ 10$^{15}$ cm$^{-2}$ in the gas phase, which is compatible with the Gibb et 
al. result to within errors; we therefore conclude that any OCS near IRS 9 is firmly sequestered as an ice.  \citet{Tielens} cites an OCS ice 
abundance of 0.1\% relative to H$_2$O ice toward IRS 9 given a water ice column density of 1 $\times$ 10$^{19}$ cm$^{-2}$.  That figure 
may be compared to a CO ice abundance of 10\%.  In interstellar ices, OCS is commonly found in association with hydrogenated species 
such as H$_2$O and CH$_3$OH (\citealt{OCS97}; \citealt{Dartois99}).  Laboratory results reported by \citet{Collings04} indicate that OCS in 
an H$_2$O mixture desorbs at $\sim$150 K, implying that significant quantities of this molecule may be frozen out in cold envelopes.  
However, C$_2$H$_2$ is seen toward both IRS 1 and IRS 9, for example, and according to the same study desorbs at only a slightly lower 
temperature.  The relative columns of OCS and C$_2$H$_2$ may then reflect real differences in their abundances rather than their 
desorption temperatures.   Alternately, OCS might only be abundant in the colder, outer envelope of IRS 9, explaining its apparent absence 
in the gas phase.  \citet{Claudia} did not observe OCS in the gas phase toward IRS 1 and \citet{Gibb04} only cite an upper limit of 1 $\times$ 
10$^{15}$ cm$^{-2}$ for the column density in the solid phase.  Other sulfur-bearing molecules desorb from grains at a relatively low 
temperature of $\sim$70 K, well below the temperatures in both objects.  It is difficult, then, to understand why high-$J$ lines of CS were 
observed toward IRS 1 by \citet{Claudia} when recent models by \citet{Wakelam11} \it underpredict \rm the abundance of CS relative to 
other S-bearing molecules at temperatures above 100 K.  The IRS 1 CS gas-phase column density is comparable to the solid-phase column 
density of OCS toward IRS 9, so if IRS 1 began with a comparable OCS abundance it may indicate a reaction route preference for forming 
CS as opposed to, e.g., SO$_2$.  However, \citet{Wakelam04} do not associate OCS and CS but rather assume CS exists in the gas of the 
pre-evaporative phase.  We did not search for CS or SO$_2$ toward IRS 9 on the basis of their presumed low abundances, and thus are 
unable to comment on the likelihood of any OCS/CS association.

The occurrence of isocyanic acid (HNCO) may be tied to the presence of the cyanate ion (OCN$^{-}$), where OCN$^{-}$ is formed on grains 
by irradiation of solid HNCO in ices by UV in the presence of NH$_3$ \citep{vB04}.  Once evaporated from grains, OCN$^{-}$ may then 
undergo hydrogenation to become HNCO again.  We speculate, however, that the hydrogenation of OCN$^{-}$ takes place \it before \rm 
desorption, since the charge an ion induces while on a grain would result in a relatively high desorption temperature.  \citet{Claudia} 
observed dozens of HNCO lines toward IRS 1 and infer a column density of 5.4 $\times$ 10$^{15}$ cm$^{-2}$; this and the value of other 
gas parameters from their spectra led them to conclude that gas-phase HNCO may originate from evaporation of grain mantles by way of 
OCN$^{-}$.  A low column density compared to the value expected from the observed abundance of OCN$^{-}$ is interpreted as evidence 
that only a fraction of the available OCN$^{-}$ desorbed from grains becomes HNCO, the remainder forming other molecules in gas-phase 
chemistry.  Interpretation of observations is difficult, however, as Knez et al. point out that most chemical models do not currently include 
HNCO in their networks.  All of the HNCO lines in the various TEXES settings we observed were non-detections.  Our fitting program 
preferred column densities of (0.1$\pm$0.5) $\times$ 10$^{16}$ cm$^{-2}$ and (0.9$\pm$1.6) $\times$ 10$^{16}$ cm$^{-2}$ in the two 
presumed absorbing units, although a clear lack of detected lines make these values limits at best.  We cannot, therefore, confidently 
distinguish between scenarios in which HNCO has a very low abundance in all phases and those in which it is effectively locked up in ices 
and unavailable to spectroscopy at our wavelengths.

\subsection{Radiative Excitation and Radiative Transfer\label{subsec:nh3scheme}}

In order to explain the observed pattern of NH$_3$ absorption and emission in our spectra of IRS 9, we invoke a radiative transfer effect that 
emerged from our modeling efforts.  We begin with gas in a shell of thickness $t$ around the continuum source with an NH$_3$ density $n_
{NH_3}$.  The rotational levels ($J$,$K$) of the ground vibrational state are assumed to be populated thermally at temperature $T$.  Electric 
dipole transitions to and from an upper vibrational level proceed according to the selection rule that ${\Delta}K$ = 0, ${\Delta}J$ = $\pm$ 1 or 0.  
In ammonia there is also a doubling of each vibrational level into symmetric and antisymmetric ($s$,$a$) inversion states, and the 
requirement that $s$ $\leftrightarrow$ $a$.  Transitions are labeled by the inversion symmetry of the lower level.  This gives the permitted 
transitions $a$/$s$ $R$($J_u$-1,$K$), $a$/$s$ $Q$($J_u$,$K$) and $a$/$s$ $P$($J_u$+1,$K$).  We obtain absorption coefficients $\alpha$ for each of 
these transitions from GEISA.

The absorption optical depth corresponding to $\alpha$ is 
\begin{equation}
\tau(w) = \alpha N(\textrm{NH}_3) {\phi}_{w}
\end{equation}
in which $N$(NH$_3$) is the radial column density through the shell and ${\phi}_w$ is the line shape function, taken to be a Gaussian.  
Molecules are excited to the upper state along all branches at a rate
\begin{equation}
r_{up} = \displaystyle\sum_{P,Q,R} \int \frac{L_w}{hcw}\tau(w) dw
\end{equation}
per second assuming ${\tau}$ ${\ll}$ 1 through the shell, where $L_w$ is the incident continuum luminosity.  The rate at which photons are 
emitted on a given branch is determined by the rate of transitions populating the upper level and the branching ratio of the Einstein $A$ 
coefficients, e.g.,
\begin{equation}
r_{down}\textrm(P\textrm) = r_{up}\textrm(P\textrm)\frac{A_P}{{A_P}+{A_Q}+{A_R}}
\end{equation}
for the $P$ branch.  We calculate values for the  $A$ coefficients according to
\begin{equation}
A_{ul} = \frac{8{\pi}c}{{\lambda}^2}{\alpha}(T)\frac{Q_r}{g_u}e^{{E_l}/kT}
\end{equation}
where $\lambda$ is the wavelength of the transition, $Q_r$ is the rotational partition function, $g_u$ is the statistical weight of the upper 
state, and $E_l$ is the energy of the lower state.

We consider the $aP$(4,$K$), $aQ$(3,$K$), and $aR$(2,$K$) lines that go to the $J$ = 3 rotational level of the $s{\nu{_2}}$ vibrational 
level.  We use the branching ratios to compute the expected emission flux in a given line $l$ with wavenumber center $w$ from
\begin{equation}
F_l = \frac{A_l}{{A_P}+{A_Q}+{A_R}} \frac{hcw}{4{\pi}d^2} \displaystyle\sum_{P,Q,R} \frac{L_w}{hcw} N{\alpha}(T).
\label{eqn:f_line}
\end{equation} 
To calculate the net flux in a line we also need the amount of absorption.  We require the relative intensity of the 10 $\mu$m dust emission 
feature at the frequency of each of the branches in order to compute the absorption line flux.  We use mid-infrared spectra of the young stars 
DI Cep and DK Tau obtained by \citet{Hanner98} to which they fit models of optically-thin silicate dust emission.  From these spectra we 
calculate relative intensities of 0.52, 0.78 and 1.00 for the continuum at the 11.7, 10.7, and 9.7$\mu$m wavelengths of the $aP$, $aQ$, and 
$aR$ lines for which $J_u$ = 3, respectively.  The intensities are normalized to the value on the $aR$ branch as it is nearest to the peak of 
the 10 $\mu$m silicate feature.  The observed continuum flux near the $P$-branch lines, $\sim$60 Jy, is turned into a luminosity at the 
source by multiplying by 4${\pi}d^{2}$, where $d$ is the distance to NGC 7538 (2.65 kpc).  We corrected for extinction using the 10 $\mu$m 
optical depth measurement of \citet{Willner82}, which implies $A_{11.7{\mu}m}$ $\sim$ 2.2 magnitudes and $A_{9.7{\mu}m}$ $\sim$ 4.4 
magnitudes toward IRS 9.

The extinction-corrected luminosity $\sim$1.5$\times$10$^{35}$ erg s$^{-1}$ cm$^{-1}$ at the position of the $P$-branch lines is scaled to 
the other branches according the values above.  We assume throughout that the lines are sufficiently broad that they remain optically thin, 
leading to uncertainty in the covering factor.  Only the product of column density and covering factor is meaningful under such 
circumstances, so we used the sum of this product for the two NH$_3$ components rather than the sum of the NH$_3$ column densities 
themselves in calculating $F_l$.  We considered the possibility that the optically thin case might result in uncertainties propagating to the 
line fluxes, and to this end we carried out additional runs of \tt FITSPEC \rm in which we deliberately held the value of the covering factor 
fixed equal to 1.  This resulted in a variation of the product of column density and covering factor of $<$ 20$\%$ compared to allowing the 
covering factor to vary.  We therefore retained the assumption of the optically thin case, and opted to use the lower value for the product of 
column density and covering factor of 8.52 $\times$ 10$^{15}$ cm$^{-2}$ in place of $N$ in Equation~\ref{eqn:f_line}.  For ${\alpha}$($T$), 
we chose an excitation temperature of 200 K, consistent with the \tt FITSPEC \rm results for the -61.6 km s$^{-1}$ component.  Finally, the 
assumed extinction was re-introduced at the end of the calculation to realistically predict the line fluxes observed at Earth.

The line fluxes we compute are given in Table~\ref{nh3flux}.   The predicted fluxes are the differences of the strengths of the predicted 
emission and absorption components of each line and show an overall pattern of emission on the $P$ branch, weak absorption or no lines 
on the $Q$ branch, and absorption on the $R$ branch.  The $P$-branch fluxes compare favorably with the integrated line fluxes from our 
data implying that our explanation is at least plausible.  This suggests that the dust and NH$_3$ are in close proximity to one another, if not 
actually mixed.
\begin{deluxetable}{ccccc}
\tablecolumns{5}
\tablewidth{0pt}
\tablecaption{Predicted fluxes for the $aP,aQ,aR$ lines of ${\nu}_2$ transitions of NH$_3$ ($J_u$ = 3) and measured fluxes for the corresponding $aP$-branch lines from TEXES spectra.}
\tablehead{
& \multicolumn{3}{c}{Predicted} & Observed \\
$K$ & F$_{aP}$ & F$_{aQ}$ & F$_{aR}$ & F$_{aP}$ \\
\hline}
\startdata
0	&	5.55	&	--	&	-1.94	& 	1.80	\\
1	&	2.64	&	0.03	&	-0.95	& 	1.35	\\
2	&	2.03	&	0.02	&	-0.73	& 	--	\\
3	&	2.11	&	-0.87	&	--	& 	0.90	\\
\enddata
\tablecomments{All fluxes are given in units of 10$^{14}$ erg s$^{-1}$ cm$^{-2}$.  Positive flux values indicate net emission while negative values indicate net absorption in a given line.}
\label{nh3flux}
\end{deluxetable} 

In the case of acetylene, we observed lines of the $\nu_5$ $Q$ and $R$ branches at 12.5-13.7 $\mu$m and the $\nu_4+\nu_5$ $P$ branch 
near 7.5 $\mu$m.  The $\nu_5$ $Q$-branch and $\nu_4+\nu_5$ $P$-branch lines are of low S/N, due to telluric interference, but clearly 
lead to higher C$_2$H$_2$ column densities than the $\nu_5$ $R$-branch lines when the observations are interpreted with the pure 
absorption model.  The $\nu_4+\nu_5$ lines appear to require at least an order of magnitude more C$_2$H$_2$ than do the $\nu_5$ R-
branch lines.  The radiative transfer effects discussed for NH$_3$ may also be relevant for C$_2$H$_2$, and may explain a part of the 
discrepancy between the C$_2$H$_2$ bands.  But the bands are farther from the silicate feature and the wavelength separation between 
the branches of the C$_2$H$_2$ bands is smaller than for NH$_3$, so the effects should be smaller.  The greater column density derived 
from the $\nu_4+\nu_5$ lines could instead be due to the smaller extinction at 7.5 $\mu$m allowing a greater column of gas to be probed.  
This explanation was suggested by \citet{Evans91} to explain a similar effect seen toward Orion IRc2, but it is difficult to understand how it 
could explain more than a factor $\sim 2$ discrepancy, given the ratio of extinction at the different wavelengths.

We propose a different explanation for the variation between the column densities derived from the C$_2$H$_2$ $\nu_5$  and $\nu_4+
\nu_5$ lines.  Vibrational excitation caused by a $\nu_5$ band photon is almost always followed by emission of another $\nu_5$ band 
photon, although possibly in a different branch.  (Due to the C$_2$H$_2$ selection rules, $P$- or $R$-branch absorption can be followed 
by either $P$- or $R$-branch emission, but $Q$-branch absorption must be followed by emission in the same branch.)  However, 7.5 $\mu$m $\nu_4+\nu_5$ absorption is most often followed by a $\nu_4+\nu_5-\nu_4$ transition at $\sim$ 13.7 $\mu$m to the $\nu_4$ level.  
Since radiative decay from the $\nu_4$ level is forbidden, the $\nu_4$ excitation energy is then eventually lost in a collision.  The result of 
this series of transitions is that there is little $\nu_4+\nu_5$ emission canceling the $\nu_4+\nu_5$ absorption, and these lines are probably 
reasonably well treated with the pure absorption model that we used.  In contrast, the $\nu_5$ absorption lines may be largely cancelled by 
emission, leading to an underestimate of the C$_2$H$_2$ column density.  The amount of re-emission into our beam is very dependent on 
the (unknown) source geometry, which determines whether our line of sight has more or less than the average for the gas surrounding the 
source.  Assuming the $\nu_4+\nu_5$ lines can be treated with a pure absorption model to give a fair measure of the abundance along our 
line of sight, the required correction to the column densities derived from the $\nu_5$ lines is approximately a factor of 10.  The required 
correction to the column densities of our other observed molecules may be similar.

\subsection{Abundance Trends and Relative Ages}

\begin{figure*}\centering
\includegraphics[angle=0,width=0.7\textwidth]{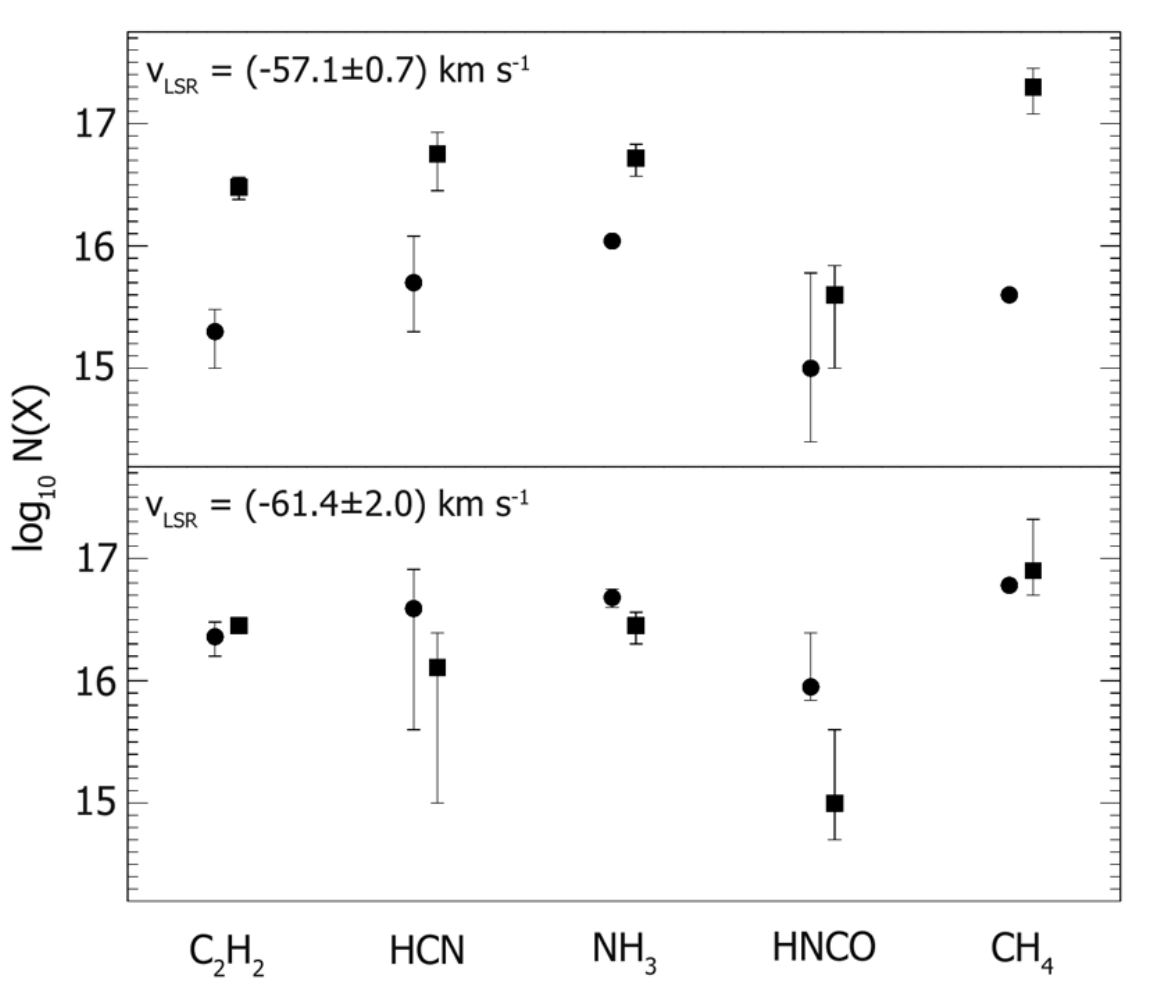}
\caption{Log column densities for the main molecular species detected in this study for the two principal velocity components toward each 
target.  IRS 9 values (this work) are shown as filled circles and IRS 1 values from \citet{Claudia} as filled squares.  The symbols for a given 
molecule are slightly offset for clarity.}
\label{column_plot}  
\end{figure*}
In Figure~\ref{column_plot} we present a graphical comparison of the column densities of C$_2$H$_2$, HCN, NH$_3$, HNCO, and CH$_4$ we obtain toward IRS 9 with those derived for IRS 1 in \citet{Claudia}.  Column densities for each of the two components with similar 
LSR velocities (-57 km s$^{-1}$ and -61 km s$^{-1}$) toward each object are shown.  We reach rather different conclusions for the two 
velocity components; IRS 1 appears to be clearly enhanced in all molecules except for HNCO at -57 km s$^{-1}$, whereas the column 
densities are essentially the same to within errors at -61 km s$^{-1}$.  HNCO is the exception at the higher negative velocity, but we 
again stress that the cited values for IRS 9 are those preferred by our fitting program despite HNCO being an evident non-detection in our 
data.  It is therefore not strictly reliable in comparison with the firm detection of HNCO toward IRS 1.  Also, we do not have any evidence \it a 
priori \rm that these two velocity components are kinematically related.   \citet{Claudia} claim the two components seen toward IRS 1 are real 
but may not be separate; they may instead represent a non-Gaussian velocity distribution.  If the latter situation is true, then such a velocity 
distribution may be systemic in other parts of NGC 7538.  Regardless, the pattern that emerges from Figure~\ref{column_plot} is suggestive 
of real differences in the gas-phase abundance of these molecules, which we attribute to differing thermal conditions in the environment of 
each object.  If the molecular abundances in all phases were the same to begin with, the gas-phase abundances we see at the present time 
are indicative of a difference in the relative ages of IRS 9 and IRS 1.

Several other lines of evidence also argue in favor of a younger age for IRS 9.  First, a number of molecules (e.g., H$_2$O, CH$_3$OH, 
HCOOH, and OCN$^{-}$) are seen in the solid phase in the spectrum of IRS 9 that are either in the gas phase or apparently absent toward 
IRS 1.  CO and CH$_4$ are seen both in the solid and gas phases, toward IRS 9 although the gas-phase lines probably form in the shocked 
outflow material.  Second, a fairly strong ionizing radiation field exists near IRS 1, whereas our non-detection of the 12.8 $\mu$m line of [\ion
{Ne}{2}] discussed in Section~\ref{subsec:neflux} argues for a much weaker field toward IRS 9.  In addition, IRS 1 has an ionized component 
to its outflow, probably driven by a wind; IRS 9 shows no similar outflow component.  Third, the chemistry of IRS 1 resembles the ``hot 
core'' phase in which ices desorb from dust grains, leading to a short-lived ($\sim$ 10$^4$-10$^5$ yr) high-temperature gas-phase 
chemistry and the synthesis of complex organic molecular species \citep{vdt05}.  Some species, such as HNCO, are clear detections toward 
IRS 1 but appear to be absent toward IRS 9; we cannot rule out an intrinsic abundance effect in explaining this difference, but a simpler 
explanation is that the IRS 9 core is not yet sufficiently warm to evaporate it.

\citet{Claudia} claim an age of $\sim$10$^5$ yr for IRS 1 based on their chemical models.  However, there is significant uncertainty in 
that value.  Abundances of the molecules they detected indicate an age closer to $\sim$10$^4$ yr, but for other molecules the chemical 
model predictions disagree with the observed values, suggesting ages in the range of 2 $\times$ 10$^3$ to 2 $\times$ 10$^6$ yr.  
They explain this large range as a result of insufficiently detailed models.  The models of \cite{Doty02} predict enhancements in the 
abundances of C$_2$H$_2$, HCN, and CH$_4$ for temperatures above about 800 K at late times ($\geq$ 10$^5$ yr); the 
enhancements are similar to those observed toward IRS 1, but the temperatures derived by \citet{Claudia} are considerably lower than 
this.  The fractional abundance of HCN in the Doty et al. model for 200 K quickly turns down at times $>$ 10$^5$ yr whereas those of C$_2$H$_2$ and CH$_4$ increase through at least 10$^{5.5}$ yr; $N$(HCN) is in fact slightly higher toward IRS 1 compared to IRS 9 by a 
factor of about 1.4.  This may both support the argument that IRS 1 is the older object of the two and suggest an upper limit for its age, since 
the HCN abundance has not begun the predicted decline, but this conclusion is tentative since our value of $N$(HCN) for IRS 9 is 
determined by the singular detection of the ${\nu}_2$ $R$(16) line.  Lastly, an additional age constraint from chemistry is provided by the 
non-detection of the molecule CCS by \citet{HoffmanKim12b} in EVLA data, suggesting an age $\geq$ 10$^4$ yr.  Given that we find IRS 9 
to be in an evidently more primitive chemical state than IRS 1, the upper limit to its age is probably $\sim$ 10$^4$ yr.  Detailed, robust 
chemical modeling, using our observations as inputs, may help more firmly constrain this value.

Our results, when considered in the context of the earlier work on IRS 1, qualitatively support a picture of an evolutionary sequence of 
cores in NGC 7538 consistent with the models of \citet{ElmegreenLada77} and \citet{CampbellThompson84} in which a wave of shock/
ionization fronts from a previous generation of stars caused the sequential collapse of the cores that became IRS 1 and IRS 9, respectively.  
As the wave continues to propagate toward the southeast, it should initiate the collapse of cores found in progressively more primitive states.  
The `IR-quiet' protostellar object NGC 7538 S is likely the next youngest object in the region after IRS 9 \citep{Pestalozzi06}.  Such a 
sequence, including NGC 7538 IRS 11 among the youngest objects, is supported by the maser observations of \citet{Hutawarakorn03}.  It is 
tempting to assume that the wave has left the IRS 1 region already, but several newly-discovered submillimeter sources within 0.35 pc of 
IRS 1 reported by \citet{Qiu11}, presumably harboring forming intermediate- or high-mass stars, may be younger than IRS 9 and 
contemporaneous with NGC 7538 S.  However, they may also be fragments, along with IRS 1, of a common progenitor.  The star formation 
history of NGC 7538 may therefore be more complex than can be represented with a simple, monotonically-increasing series of protostellar 
ages.

\subsection{Spatial Resolution of IRS 9\label{subsec:scanmap}}

We checked whether IRS 9 is spatially resolved in TEXES data by obtaining spatial-spectral maps of IRS 9 and $\mu$ Cep, a mid-infrared-
bright standard star that was used as a PSF reference for spatial deconvolution.  To generate the maps, the TEXES slit was stepped across 
each object and spectral information gathered at each step, building up data cubes in which there are two spatial dimensions and one 
spectral dimension.  Additional steps were included at both the beginning and end of the scans to sample the sky for removal.  The data are 
handled by our reduction pipeline in a manner substantially similar to the method of processing the nodded observations, treating a scan as 
a series of 2-D spectrograms.  

\begin{figure}
\leavevmode
\includegraphics[width=0.5\textwidth]{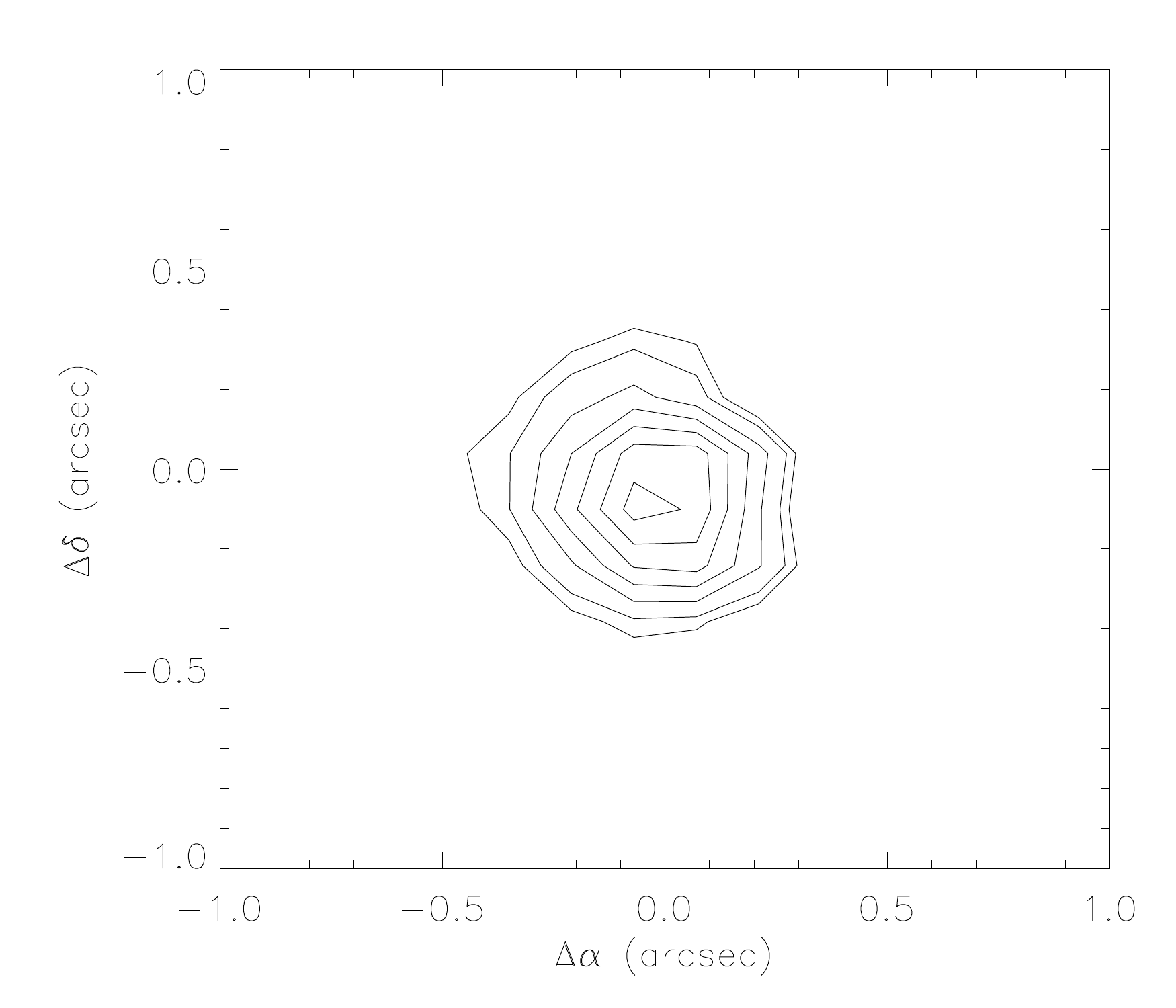}
\caption{Contour plot of the TEXES scan map of NGC 7538 IRS 9 at 745.8 cm$^{-1}$ after maximum entropy deconvolution.  The PSF 
reference for the deconvolution was a scan map of the mid-infrared standard $\mu$ Cep made with the same parameters as the IRS 9 scan.  
North is up and east at left and contours are plotted at flux densities of 25, 50, 100, 150, 200, 250 and 300 Jy.  The plot origin is 
approximately the location of the source peak intensity.}
\label{irs9_deconv}
\end{figure}

IRS 9 and $\mu$ Cep were observed with identical scan parameters including scan direction, length, and step size, and were observed at 
comparable zenith angles.  After reducing the raw data, different spectral windows were selected from which to form reconstructed images of 
each object by summing the data in the spectral direction.  For IRS 9, we chose a window centered at about 745.88 cm$^{-1}$, near the 
position of the C$_2$H$_2$ ${\nu}_5$ $R$(6) line in the rest frame of the object.  The window was 0.35 cm$^{-1}$ wide.  We selected this 
frequency range to search for spatial extent in the IRS 9 map as we noted absorption in the $R$(6) line in nod-mode spectra.  For the 
deconvolution reference, we chose a nearby spectral region apparently free from telluric and stellar photospheric absorption centered at 
744.43 cm$^{-1}$ and 0.25 cm$^{-1}$ in width.  The observed FWHM for each object was 0$\farcs$67 for IRS 9 and 0$\farcs$63 for $\mu$ 
Cep at each object's respective spectral setting.  The deconvolution was performed using the maximum entropy method (\citealt{Bryan80}; 
\citealt{Narayan86}), with results shown in Figure~\ref{irs9_deconv}.  The deconvolved image of IRS 9 in the C$_2$H$_2$ feature appears 
pointlike with some possible weak extension toward the NE.  We do not believe this indicates that the envelope or other structure of IRS 9 is 
spatially resolved,  and we carried out our other TEXES observations exclusively in nod mode.

This approach to imaging and deconvolution should be interpreted cautiously.  Poor seeing during a scan, for example, might distort the 
PSF on time scales shorter than the time required to scan across an object.  However, we felt sufficiently convinced that IRS 9 presents itself 
as a point source in the mid-infrared that we did no further scans.  Given the distance to NGC 7538 and the spatial resolution of our 
observations, we conclude that spatial structure of IRS 9 is unresolved on scales $\leq$ 2000 AU.  This limit is consistent with the 
inferred sizes of disks detected around protostars with 10$^3$ - 10$^5$ $L_{\odot}$ (\citealt 
{Patel05,Schreyer06,Rodriguez07,FH09,GM10}) and places a constraint on physical models of massive protostars in NGC 7538, which 
often include disks thought to be hundreds of AU in size (e.g., \citealt{Sandell09,Surcis11}).

\subsection{A Structural Model Of IRS 9\label{subsec:toymodel}}

\begin{figure}
\leavevmode
\includegraphics[width=0.5\textwidth]{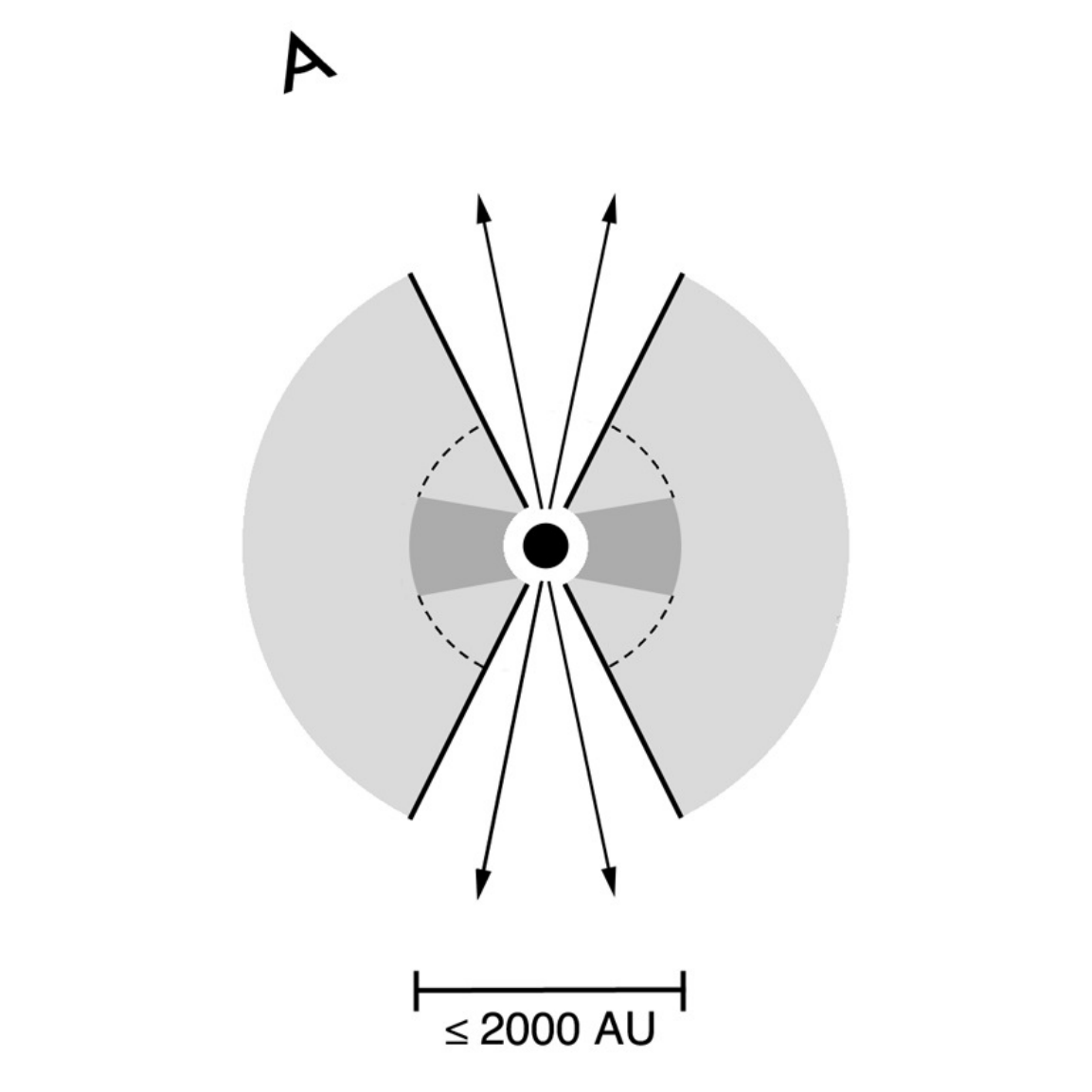}
\caption{A cartoon rendering for a proposed model of NGC 7538 IRS 9 based on TEXES data.  A central, embedded source (or sources) has 
at least partially cleared cavities in a bipolar outflow while significant dust remains in a spherical distribution (light gray shading).  An inner 
disk may exist (dark gray shading).  To account for our observations, the line of sight samples both scattered light in the outflow as well as a 
``dust photosphere'' (dashed lines) where the dust becomes opaque at a mid-infrared wavelengths.  The inferred viewing angle is shown, 
and a scale bar is provided based on the constraints discussed in Section~\ref{subsec:scanmap}.  Note that the envelope and outflow 
cavities as depicted are not shown to this scale.}
\label{toymodel}
\end{figure}
Based on the conclusions we reach from kinematic indications in the data, we compose a toy model of the IRS 9 system to show its essential 
structure and a guess at the viewing geometry.  This model is shown in Figure~\ref{toymodel}.  In this view, the central, perhaps accreting 
object is shown in cross-section with a surrounding disk.  The known bipolar outflow is indicated by the evacuated regions above and below 
the disk and by arrows showing the direction of the flow.  A lighter shading above and below the disk represents the remaining envelope 
that has not been accreted or become part of the outflow.  In order to account for seeing certain molecules under presumed LTE conditions 
while not observing the exciting source directly, we suggest a ``dust photosphere'' at a radius from the source at which the emission from the 
dust becomes optically thick, as indicated by dashed lines in the figure.  It is this effective source of radiation against which we see most of 
our absorption lines.  Since we do not resolve spatial structure in the molecular absorption region above the dust photosphere, we interpret 
the size scale constraint discussed in the previous section (2000 AU) as an upper limit to the radius of the dust photosphere itself.  
This value also represents an upper limit of the size of the disk, because a larger, optically-thick disk should have been detectable by virtue 
of its spatial extent.  Our data do not permit a constraint to be placed on the size of the surrounding envelope or the outflow cavities.  

The inclusion of a disk in this picture is tentative, given that the lifetime of disks around massive stars is 
thought to be particularly short, of order 10$^{-4}$ the lifetime of the host star \citep{MannWilliams}.  The evolutionary state of IRS 9 may be 
sufficiently early in its history, well before the embedded protostar is visible, to retain a disk.  \citet{Claudia} argue for a disk around IRS 1 
whose surface is observed at near-grazing incidence; a sightline through the disk atmosphere is proposed to explain their observations.  

We propose a viewing angle based on the following.  First, the broad CO lines with their large, Doppler-shifted velocity separations suggest 
formation in the bipolar outflow.  Second, the apparent radiative ``pumping'' of the NH$_3$ transitions, discussed in Section~\ref
{subsec:nh3scheme}, requires continuum light from the direction of the dust photosphere.  Third, the other molecular species we observed 
in absorption also require a background continuum source, though the range of excitation conditions we derive is insufficient to definitively 
locate them in the envelope or the turbulent  edge of the outflow cavity.  The C$_2$H$_2$ combination mode may be an exception and 
could shock-excited in the outflow.  Consequently, we place the viewing angle such that the observer is looking nearly along the edge of the 
outflow cavity and sees light from both sources.  This model appears to account for all aspects of our observations in at least a qualitative 
way.  It is also consistent with the conclusion of \citet{Sandell05}, based on HCO$^+$ $J$ = 2$\rightarrow$1 observations showing deep, 
redshifted self-absorption, that the molecular outflow is viewed nearly pole-on.

\section{Summary\label{sec:summary}}

We have presented the method and results of a study of the embedded high-mass protostellar object NGC 7538 IRS 9 and compared our 
findings to a similar investigation of the related object NGC 7538 IRS 1.  We obtained high resolution, mid-infrared spectra of these objects 
in 46 ro-vibrational transitions of the fundamental bands of the molecules C$_2$H$_2$, CH$_4$, HCN, NH$_3$ and CO and a number of 
their isotopologues.  We also detected two lines of the ${\nu}_4+{\nu}_5$ combination mode of C$_2$H$_2$. From these observations we 
draw some broad conclusions.

1. IRS 9 appears to be a spatially-unresolved object on a scale of $\sim$2000 AU.

2. With the exception of CO, whose lines are saturated, we did not observe $^{13}$C isotopologues of any organic molecules.    This may 
indicate a particularly large value of the $^{12}$C/$^{13}$C ratio, or the $^{13}$C isotopologue lines were simply below our threshold of 
detection.

3. There is no discernible, consistent trend in the abundance variations of C$_2$H$_2$, HCN, NH$_3$, and CH$_4$ with respect to either 
CO or H$_2$ between IRS 9 and IRS 1.  Column density variations between the two objects are an order of magnitude or less in each case. 

4. The observation of gas-phase CS absorption toward IRS 1 and the evident non-detection of sulfur-bearing species like OCS in the gas 
phase toward IRS 9 may reflect a real S abundance variation between the two objects but can also be explained by different thermal 
conditions in each case.  A similar abundance discrepancy involving HNCO may indicate sufficiently low temperatures in the IRS 9 
envelope to retain it as an ice and explain its detection in the solid phase by \citet{Gibb04}.  If the abundances in all phases are actually 
comparable between the two sources, a relatively weak heating source is implied in the case of IRS 9.  While this is at odds with the 
somewhat higher inferred abundances of the other molecules toward the warmer environment of IRS 1, the suggested sightline in that 
direction through a disk atmosphere subjects those molecules to direct irradiation by protostellar UV.  This may drive photochemistry that 
lowers abundances of certain molecules as they are converted to other kinds.

5. However, the relative abundances of various molecular species with respect to CO and H$_2$ can also be explained by variations in the 
CO/H$_2$ ratio or the gas-to-dust ratio between the two objects.  Choosing between these interpretations would require better determining 
the CO and H$_2$ abundances toward both objects.

6. Lines of many species observed toward IRS 9 have higher Doppler $b$ values than those toward IRS 1, resulting in lines toward IRS 9 
being less saturated than those toward IRS 1.

7. Given the non-detection of \ion{Ne}{2} in our data and the continuum flux density at the expected wavenumber of the line (70$\pm$10 Jy), 
we find a corresponding upper limit of $\sim$10$^{45}$ photons s$^{-1}$ for the Lyman continuum flux in IRS 9.  This value is well below 
both the expected flux of a ZAMS star with the same luminosity and the ionizing flux of IRS 1 ($>$ 10$^{48}$ photons s$^{-1}$).

8. The observed pattern of emission and absorption components of ammonia lines in our spectra can be explained by a non-LTE radiative 
transfer effect involving ``pumping'' of lines on the $R$-branch by light from the 9.7 $\mu$m silicate dust feature.  Predictions of the strength 
of this effect compare reasonably well with observed fluxes in $P$-branch lines.  

9. Our data are consistent with a simple model of the IRS 9 system in which the sightline probes both a high-speed outflow and a quiescent 
envelope illuminated from behind by a ``dust photosphere'' such that line formation depth is wavelength-dependent.  The possibility of 
seeing the turbulent flow along the wall of the outflow cavity is supported by the broad, blueshifted wing of the C$_2$H$_2$ ${\nu}_{4}+{\nu}_{5}$ combination band that may arise in a shock.  A disk is included in the model given the system's inferred young age and the presence of a kinematically-indicated disk toward the (presumably) older IRS 1.

10. Observed differences in the abundance of various molecular species toward IRS 9 and IRS 1 imply an upper limit to the age of IRS 9 of $\sim$ 10$^4$ yr, whereas IRS 1 may be an order of magnitude older.  This conclusion is consistent with the triggered star formation models 
of \citet{ElmegreenLada77} as applied to NGC 7538 by \citet{CampbellThompson84} and others.

\acknowledgements{The authors wish to thank Marty Bitner, Dan Jaffe, Tommy Greathouse, and Matt Richter for assistance in obtaining the 
observations presented here.  We recognize the efforts of the support staff at both the IRTF and Gemini who helped make our observing 
runs productive and efficient.  We also gratefully acknowledge the constructive comments and suggestions of an anonymous referee, which 
substantially improved the manuscript before publication.  The data presented here are based in part on observations obtained at the 
Gemini Observatory, which is operated by the Association of Universities for Research in Astronomy, Inc., under a cooperative agreement 
with the NSF on behalf of the Gemini partnership: the National Science Foundation (United States), the Science and Technology Facilities 
Council (United Kingdom), the National Research Council (Canada), CONICYT (Chile), the Australian Research Council (Australia), CNPq 
(Brazil) and SECYT (Argentina).  Additional data were obtained at the Infrared Telescope Facility, which is operated by the University of 
Hawaii under Cooperative Agreement no. NNX08AE38A with the National Aeronautics and Space Administration, Science Mission 
Directorate, Planetary Astronomy Program.  JCB and JHL acknowledge the support of the National Science Foundation (NSF) through grant 
AST-0607312.}

\end{document}